\documentstyle[preprint,aps,tighten]{revtex}

\def\pol{\vec}
\def\half{\frac{1}{2}}

\begin{document}
\draft
\preprint{IUCM94-014}

\title{Introduction to the Physics of the Quantum Hall Regime}
\author{A.H. MacDonald}
\address{Indiana University, Bloomington, IN 47405, USA }
\maketitle
\begin{abstract}
These lecture notes attempt to explain the  main ideas of the theory of
the quantum Hall effect.  The emphasis is on the localization and
interaction physics in the extreme quantum limit which gives rise to the
quantum Hall effect.   The interaction physics in the extreme quantum limit
which is responsible for the fractional quantum Hall effect is discussed
at length and from an elementary point of view.
\end{abstract}

\narrowtext

\section{What is the Quantum Hall Effect?}

The quantum Hall effect occurs in two-dimensional electron systems in the
limit of strong perpendicular magnetic fields. Two-dimensional electron
gas systems may be realized at the interface between semiconductors and
insulators or at the interface between two different semiconductors as
illustrated in Fig.~[\ref{fig:2deg}].
The theory of the quantum Hall effect now takes as its subject all
physical properties of two-dimensional electron systems in the limit where
the magnetic field is so strong that the mixing of Landau levels by
disorder or by electron-electron interactions may be considered as a weak
perturbation. I will refer to this limit as the {\em quantum Hall regime.}
Qualitatively~\cite{disorderpot}, physical properties then depend only on
the ratio of the disorder potential to the electron-electron interaction
potential. These systems have unusual and interesting properties because
there is no energy scale, like the band width of a periodic solid or the
Fermi energy of an electron liquid in the absence of a magnetic field,
which is associated with a simple one-body term in the Hamiltonian and can
be used as the basis of a perturbation theory. Disorder and
interactions, and in the general case both disorder {\em and} interactions
have to be accounted for using some non-perturbative approach. Neither
interactions nor disorder can ever be considered to be weak. The ratio
between the energy scales corresponding to these two interactions is,
however, important. The limit where the disorder potential is much
stronger, is referred to as the {\em integer} quantum Hall regime whereas
the limit where the interaction potential is much stronger, is referred to
as the {\em fractional} quantum Hall regime. (A given sample may be in
the integer quantum Hall regime at one magnetic field strength and in the
fractional quantum Hall regime at another magnetic field strength.)

One of the challenges in probing the properties of a two-dimensional
electronic systems experimentally is in isolating it
from its three-dimensional semiconducting or insulating
host. This isolation is most easily achieved in measurements of the
electrical transport properties and
this type of measurement has been the most widely used technique for
studying two-dimensional electron systems. The quantum Hall effect was
discovered~\cite{qhedisc} by Klaus von Klitzing, nearly fifteen years ago
now, while performing measurements of the electrical transport properties
of a two-dimensional electron gas system at the strong magnetic field
facility in Grenoble. The experimental setup for these measurements is
illustrated schematically in Fig.~[\ref{fig:hallbar}].
At weak magnetic fields the magnetotransport properties of two-dimensional
electron gas are well described by the simple Drude theory, outlined in
the following section. According to this theory the dissipative resistance
$R$ of the two-dimensional electron gas system should be proportional to
$n^{-1}$ where $n$ is the areal density of the two-dimensional electron
system, while the Hall resistance $R_H$ should be proportional to $B/n$.
What von Klitzing saw was dramatically different. His discovery signaled
the occurrence of novel physical phenomena in the quantum Hall regime and
engendered a large body of work which has led to a fairly complete
understanding of many phenomena which had not even been anticipated prior to
his experiments.  These notes attempt to review a portion of that new
knowledge.

\section{Drude Theory of Magnetotransport}

Before discussing what von Klitzing actually observed we briefly review
the Drude theory~\cite{ashcroftmermin} of magnetotransport in
two-dimensions. The Drude theory gives results which are qualitatively
correct outside of the quantum Hall regime. In the simplest
version of this theory it is assumed
that the electrons of the two-dimensional electron gas are accelerated by
external forces between scattering events to reach a drift velocity,
\begin{eqnarray}
\pol{v}_D &=& \frac{\pol{F}\tau}{m^\ast}\nonumber\\
&=& -\frac{e\tau}{m^\ast}\;\left[\pol{E} + \frac{\pol{v}_D}{c}\; B\times
\hat{z}\right]
\end{eqnarray}
where $\pol{F}$ is the force experienced by the electrons, $m^\ast$ is
the effective mass of the electrons and $\tau$ is the time between
scattering events. The magnetic field enters this theory through the
appearance of the Lorentz force on the moving electrons. In the Drude
theory it is assumed that all electrons drift together so that the current
density is related to the electronic drift velocity by
\begin{equation}
\pol{j} = -ne\;\pol{v}_D.
\end{equation}
Given these equations it is possible to solve for the electric fields
present in the two-dimensional electron system given a uniform current
density:
\begin{equation}
\pol{E} = \frac{m^\ast}{ne^2\tau}\;
\pol{j} + \frac{B}{nec}\; \hat{z}\times \pol{j}
\end{equation}
The dissipative resistivity $\rho_{xx}$ (or for a square sample in
two-dimensions the resistance $R$) is given by the ratio of the electric
field in the direction of current flow to the current density while the
Hall resistivity $\rho_{xy}$ (and the Hall resistance $R_H$) is given by
the ratio of the electric-field component perpendicular to the direction
of current flow to the current density. Notice that in the Drude theory
$\rho_{xx}$ is independent of magnetic field while $\rho_{xy}$ is
proportional to magnetic field. Similar results are obtained with more
sophisticated semiclassical theories of electrical
transport~\cite{ashcroftmermin}, with the added benefit that explicit
expressions capable of quantitative accuracy are obtained for the
scattering time, $\tau$.

These theoretical results may be compared with typical experimental
results obtained in the quantum Hall regime which are illustrated in
Fig.~[\ref{fig:intqhe}] and Fig.~[\ref{fig:fracqhe}].
For fields below $\sim 0.1 {\rm Tesla}$ the Drude theory works well. At
stronger fields, however, the Hall resistivity becomes nearly constant over
certain finite intervals of magnetic field. Over the same magnetic field
intervals the dissipative resistivity becomes very small. These features
are labeled in Fig.~[\ref{fig:fracqhe}] by a set of integers and
fractions with odd denominators which we will presently see correspond to
Landau level filling factors. In what follows I will attempt to explain
why these anomalies in the transport properties of two-dimensional
electron systems occur. We'll find that the quantum Hall effect reflects
both novel disorder physics and novel interaction physics in the quantum
Hall regime.

\section{Free 2D Electron in a Magnetic Field}

Since the quantum Hall regime is defined by Landau quantization of the
kinetic energy, any discussion of the quantum Hall effect must begin with
the quantum mechanics of a single free-particle moving in two-dimensions
in a perpendicular magnetic field.  We'll find it useful to discuss the
motion of classical particles in a magnetic field first.

\subsection{Classical Solution}

I start with the solution of the classical equations of motion. Here and
throughout much of these notes it is convenient to use a complex number
notation for two-dimensional vectors. In particular we'll let $z = x + iy$
represent the two-dimensional position vector and the complex number $v =
v_x + iv_y$ represent the two-dimensional velocity vector. The classical
equations of motion take a compact form when the complex number notation
is used:
\begin{equation}
\left.\begin{array}{c}
m\ddot{x} = -\frac{eB}{c}\;\dot{y}\\
m\ddot{y} = \frac{eB}{c}\;\dot{x}\end{array}\right\rbrace\; \ddot{z} =
i\omega_c\dot{z}
\end{equation}
Integrating twice we obtain
\begin{eqnarray}
\dot{z} &=& v_0e^{i\omega_ct}\\
z &=& C - \frac{iv_0e^{i\omega_ct}}{\omega_c}
\label{eq:classcyclotron}
\end{eqnarray}
where $v_0$ is the complex number representing the velocity at time $t =
0$. Classical particles moving in two-dimensions in the presence of a
perpendicular magnetic field execute circular (cyclotron) motion with an
angular frequency $\omega_c = e B / m^* c$. The tangential velocity $v_c$
and the radius for the cyclotron orbits are related by $R_c = v_c /
\omega_c$. In Eq.~(\ref{eq:classcyclotron}) $C$ is a complex integration
constant which specifies the position vector for the center of the cyclotron
orbit.  Cyclotron orbit motion is illustrated in Fig.~[\ref{fig:cyorb}].

\subsection{Quantum Solution}

The Hamiltonian for an electron moving in two-dimensions in a perpendicular
magnetic field is given by
\begin{equation}
H = {\pol{\pi}^2 \over 2 m^*}
\end{equation}
where the kinetic momentum is given in a coordinate representation by
\begin{equation}
\pol{\pi} = - i \hbar \pol{\nabla} +{ e \pol{A} \over c}.
\end{equation}
For a uniform magnetic field $\hat z \cdot (\pol{\nabla} \times \pol{A}) =
B$ so that the vector potential $\pol{A}$ is a linear function of the
spatial coordinates. It follows that $H$ is a generalized
two-dimensional harmonic oscillator Hamiltonian which is quadratic in both
the spatial coordinates and in the canonical momentum $\pol{p} = - i
\hbar \pol{\nabla}$. The eigenstates and eigenvalues of this Hamiltonian
may be obtained by a convenient algebraic method, analogous to the `ladder
operator' solution of the one-dimensional harmonic oscillator. The
algebraic solution facilitates many calculations we will outline later in
these notes. I start from the observation that the $x$ and $y$
components of the kinetic momentum are canonically conjugate coordinates:
\begin{equation}
[\pi_{x}, \pi_y] = \frac{-i\hbar e}{c}\;\hat{z}\cdot
(\pol{\nabla}\times\pol{A}) = \frac{-i\hbar^2}{\ell^2}
\end{equation}
where $\ell^2 = \hbar c / eB$. $\ell$ is known as the magnetic length
and is the natural length unit in the quantum Hall regime. In these notes
we will always take $\ell$ as the unit of length although it will
frequently be exhibited explicitly for clarity. Note that $\ell$ is
related to the magnetic flux quantum by
\begin{equation}
2\pi\ell^2B = \Phi_0
\end{equation}
I define the first set of ladder operators as follows:
\begin{equation}
a^\dagger \equiv \frac{\ell /\hbar}{\sqrt{2}} (\pi_x + i\pi_y)
\end{equation}
so that
\begin{equation}
 [a, a^\dagger ] = 1
\end{equation}
and
\begin{equation}
H = \frac{\hbar\omega_c}{2} (aa^\dagger + a^\dagger a).
\label{eq:kineticeng}
\end{equation}
It follows from Eq.~(\ref{eq:kineticeng}) that the eigenenergies for the
Schr\"{o}dinger equation of a free particle are $\hbar \omega_c (n+1/2)$.
However, just as the classical kinetic energy is independent of the
center coordinate for the cyclotron orbit we might expect that each of
these quantum eigenenergies will be degenerate. The degeneracy is
revealed by constructing ladder operators from the quantum orbit-center
operators:
\begin{equation}
C = z + \frac{i \pi }{m^ * \omega_c}.
\end{equation}
I first note that the $x$ and $y$ components of the cyclotron orbit
centers are canonically conjugate coordinates:
\begin{equation}
[C_x, C_y] = i\ell^2.
\end{equation}
This allows us to define a ladder operator by
\begin{equation}
b \equiv \frac{1}{\sqrt{2} \ell} (C_x + iC_y)
\end{equation}
It is readily verified that
\begin{equation}
[b, b^\dagger ] = 1
\end{equation}
and that
\begin{equation}
[a, b] = [a^\dagger , b] = [H, b] = 0.
\end{equation}
The cyclotron-orbit-center ladder operators produce a set of degenerate
eigenstates of the one-body kinetic energy operator. The set of all
eigenstates with a given allowed kinetic energy is called a Landau level.
The full set of eigenstates can be generated by using raising operators
starting from the bottom of the ladder:
\begin{eqnarray}
\vert n, m\rangle &=& \frac{(a^\dagger )^n(b^\dagger )^m}{\sqrt{n!m!}} \vert
0, 0\rangle\nonumber\\
\varepsilon_n &=& \hbar\omega_c (n + \half ).
\label{eq:onebody}
\end{eqnarray}

For many calculations it is useful to choose a specific gauge for the
vector potential. In the symmetric gauge,
\begin{equation}
\pol{A} = \frac{B}{2} (-y, x, 0)
\end{equation}
and the ladder operators may be written in the form,
\begin{eqnarray}
b &=& \frac{1}{\sqrt{2}} \left(\frac{z}{2\ell} +
2\ell\frac{\partial}{\partial\bar{z}}\right)\\
b^\dagger &=& \frac{1}{\sqrt{2}} \left(\frac{\bar{z}}{2\ell} -
2\ell\frac{\partial}{\partial z}\right)\\
a^\dagger &=& \frac{i}{\sqrt{2}} \left(\frac{z}{2\ell} -
2\ell\frac{\partial}{\partial\bar{z}}\right)\\
a &=& \frac{-i}{\sqrt{2}} \left(\frac{\bar{z}}{2\ell} +
2\ell\frac{\partial}{\partial z}\right).
\label{eq:ladders}
\end{eqnarray}
(Throughout these notes we use an overbar to indicate complex conjugation.)
It follows that the orbital wavefunction at the bottom of both the
kinetic-momentum and orbit-center ladder operator chains, {\em i.e.,} the
one which is annihilated by both $a$ and $b$ is
\begin{equation}
\psi_{0,0} = \frac{1}{\sqrt{2\pi\ell^2}} e^{-z\bar{z}/4\ell^2}.
\label{eq:bottomladder}
\end{equation}
The set of single-particle orbitals in the lowest Landau level can then be
generated by repeated application of $b^{\dagger}$:
\begin{equation}
\psi_{0,m} = \frac{ \bar{z}^m
e^{-z\bar{z}/4\ell^2}}{\sqrt{2\pi\ell^2 2^m
m!}}.
\end{equation}
For the symmetric gauge it is convenient to change definitions and drop
the factors of $i$ and $-i$ which appear in the expressions for $a$ and
$a^\dagger$; this simply changes the phase convention for the higher
Landau level wavefunctions.

I will make frequent use of these symmetric gauge wavefunctions and the
ladder operator approach we used to obtain them.  I start here by deriving
an expression for the number of states per unit area in a single Landau
level. I first note that the complex coordinates for the position vector
can be expressed in terms of ladder operators as follows:
\begin{eqnarray}
z &=& \sqrt{2}\ell (b + a^\dagger )\\
\bar{z} &=& \sqrt{2}\ell (b^\dagger + a).
\end{eqnarray}
It follows that
\begin{equation}
\pi \langle n,m | \bar{z} z | n, m \rangle = 2 \pi \ell^2 (m+n+1).
\label{eq:llorbrad}
\end{equation}
For large $m$ these orbitals are strongly localized to within $\sim \ell$
of a ring with an radius $r_{m,n} = \ell \sqrt{2(m+n+1)}$ as can be verified by
calculating $\langle n,m | \bar{z}^2 z^2 |n, m \rangle$ and comparing
with the above result. When the lowest Landau level is filled by
occupying all single-particle states with $n=0$ and $m=0, \cdots , N-1$ it
follows from Eq.~(\ref{eq:llorbrad}) that in the thermodynamic limit the
area occupied per electron is $2 \pi \ell^2 $. Each Landau level contains
one single-particle state per magnetic flux quantum penetrating the area
occupied by the electrons. It is the macroscopic Landau level degeneracy
which creates the opportunity for unique physical phenomena in the strong
magnetic field limit. Another way to obtain the same result is to note
that
\begin{equation}
\sum_{m=0}^{N-1} | \psi_{0,m} (z) |^2 = (2 \pi \ell^2)^{-1}
\sum_{m=0}^{N-1} {x^m \over m!} \exp (-x) \rightarrow (2 \pi \ell^2)^{-1}
\end{equation}
with $x \equiv |z|^2/2 \ell^2$.

\subsection{Useful Identities for Symmetric-Gauge Free-Particle Eigenfunctions}

I now pause to mention and sketch the derivations of a number of
identities involving the symmetric gauge free-particle eigenfunctions
which will enter concrete calculations later in these notes. This section
of the notes can be skipped on a first reading.

\noindent{\em Plane Wave Matrix Elements}

The matrix element of a plane wave $\exp (- i \pol{k} \cdot \pol{r})$
can be evaluated by expressing the position in term of ladder
operators:
\begin{eqnarray}
&&\langle n', m'\vert e^{-i\pol{k}\cdot\pol{r}}\vert n, m\rangle\nonumber\\
&&= \langle n', m'\vert e^{-i\bar{k}(b + a^\dagger
)/\sqrt{2}} e^{-ik(b^\dagger + a)/\sqrt{2}}\vert n,m\rangle\nonumber\\
&&= e^{- \vert k\vert^2/2} \langle n'\vert e^{-i\bar{k}a^\dagger
/\sqrt{2}} e^{-ika/\sqrt{2}}\vert n\rangle
\langle m'\vert e^{-ikb^\dagger /\sqrt{2}}  e^{-i\bar k b/\sqrt{2}} | m
\rangle
\label{eq:pwme}
\end{eqnarray}
The matrix element may be written as the product of a factor associated
with the kinetic-momentum ladder operators and a factor associated with
the orbit-center ladder operators. This factorization of operators is
frequently possible in the quantum Hall regime and we will find occasion
to exploit it. Each factor may be evaluated using the algebra of the
ladder operators. For example,
\begin{eqnarray}
&&\langle m'\vert e^{-ikb^\dagger /\sqrt{2}} e^{-i\bar{k} b / \sqrt{2}} |
m \rangle \equiv G_{m',m}(k)\nonumber \\
&&= \left(\frac{m!}{m'!}\right)^{1/2}
\left(\frac{-ik}{\sqrt{2}}\right)^{m'-m} L_m^{m'-m}
\left(\frac{k\bar{k}}{2}\right)
\label{eq:projected}
\end{eqnarray}
where $L_m^{m'-m}$ is a generalized Laguerre polynomial. The second
equality in Eq.~(\ref{eq:projected}) follows (for $m' > m$) after noting
that non-zero matrix elements will arise only when the lowering operator
acts no more than $m$ times and the raising operator acts $m'-m$ times
more often than the lowering operator. The result is a sum of $m$ terms
which can be compared with the definition of the generalized Laguerre
polynomials. With this definition we have the result
\begin{equation}
\langle n', m'\vert e^{-i\pol{k}\cdot\pol{r}}\vert n, m\rangle
= \exp ( - |k|^2/ 2) G_{n',n}(\bar k) G_{m',m} (k).
\end{equation}
It is also useful to note from the above definitions that the projection
ofa plane wave onto the lowest ($n =0$) Landau level is
\begin{equation}
 e^{-i\bar{k}b /\sqrt{2}} e^{-i k b^{\dagger} /\sqrt{2}} \equiv B(k).
\label{eq:pwproj}
\end{equation}
Except for the factor of $\exp ( - | k |^2 /2) $ which comes from
interchanging the orders of the raising and lowering factors in
Eq.~(\ref{eq:pwproj}) we see that $G_{n,m} (k)$ is the
symmetric gauge single-particle
eigenstate representation of the projected plane-wave operator.

\noindent{\em Inversion}

{}From the explicit form of Eq.~(\ref{eq:projected}) it follows that
\begin{equation}
G_{m',m}(-k) = (-)^{m'-m} G_{m',m}(k)
\end{equation}

\noindent{\em Symmetric Gauge Wavefunctions}

By letting the two raising operators act repeatedly on
$\phi_{0,0}(\pol{r})$ we find that
\begin{equation}
\langle\pol{r}\vert n,m\rangle\equiv\varphi_{n,m} = \frac{e^{-\vert
z\vert^2/4}}{\sqrt{2\pi}} G_{m,n}(i\bar{z})
\label{eq:wavefunctions}
\end{equation}
The similarity of the plane-wave matrix elements and the orbital
wavefunctions is reminiscent of the position-space, momentum-space
duality which occurs for the one-dimensional harmonic oscillator.

\noindent{\em Matrix Products}
We can derive an expression for the product of two
$G_{m,m'}(k)$ matrices with different wavevector arguments by using the
completeness of the ladder operator eigenstates:
\begin{eqnarray}
\sum_\ell G_{m',\ell}(k_1) G_{\ell ,m}(k_2) &=& \sum_\ell \langle m'\vert
e^{-ik_1b^\dagger /\sqrt{2}}
e^{-i\bar{k}_1b/\sqrt{2}}\vert\ell\rangle\nonumber\\
&&\times\langle\ell\vert e^{-ik_2b^\dagger /\sqrt{2}}
e^{-i\bar{k}_2b/\sqrt{2}}\vert m\rangle\nonumber\\
&=& e^{-\bar{k}_1k_2/2} G_{m',m}(k_1 + k_2).
\label{eq:matrixproduct}
\end{eqnarray}
The additional factor in Eq.~(\ref{eq:matrixproduct}) comes from
interchanging raising and lowering factors after invoking completeness.

\noindent{\em Hermitian Conjugate}

Taking the complex conjugate of Eq.~(\ref{eq:projected}) we find
immediately that
\begin{equation}
\bar{G}_{m',m}(k) = G_{m,m'}(-k) = (-)^{m-m'} G_{m,m'} (k)
\end{equation}
This result is also obvious from the explicit expression for $G_{n,m}(k)$
in term of Laguerre polynomials.

\noindent{\em Landau Level Degeneracy}

I have previously derived a result for the number of states per unit area
in the lowest Landau level. This result can be generalized to arbitrary
Landau level using Eq.~(\ref{eq:matrixproduct}) and
Eq.~(\ref{eq:wavefunctions}):
\begin{eqnarray}
\sum_m \vert\varphi_{n,m}\vert^2 &=& \frac{e^{-\vert z\vert^2/2}}{2\pi} \sum_m
G_{n,m}(-i\bar{z}) G_{m,n}(i\bar{z})\nonumber\\
&=& \frac{1}{2\pi\ell^2} G_{n,n}(0) =
\frac{1}{2\pi\ell^2}.
\end{eqnarray}
This implies that the number of states per unit area per Landau level is
$N_{\phi} = B A / \Phi_0$, the number of flux quanta which pass through
the area of the system.

\noindent{\em Full Landau Level Rule}

The fact that the charge density of a full Landau level is a constant
gives us a result for the trace of the plane-wave matrix elements:
\begin{equation}
\sum_{m=0}^\infty G_{m,m}(k) = \int d\pol{r} e^{-i\pol{k}\cdot\pol{r}} e^{\vert
k\vert^2/2} (2\pi\ell^2)^{-1} = N_\phi\delta_{\pol{k},0}.
\label{eq:fll}
\end{equation}

\noindent{\em Orthogonality}

We will have occasion to expand quantities of physical interest in terms
of the plane-wave matrix elements. The following orthogonality
relationship will be useful:
\begin{eqnarray}
\int d^2\pol{k} e^{-\vert k\vert^2/2} G_{m',m}(k) G_{n',n}(\bar{k}) &=&
\bar{\varphi}_{n',m'}(0) \varphi_{n,m}(0)\nonumber\\
&=& \frac{\delta_{m',n'}\delta_{m,n}}{2 \pi}.
\label{eq:orthogonality}
\end{eqnarray}
This equation can be understood by recognizing the left-hand side as the
Fourier expansion of the matrix element of
$\delta (\pol{r})$ between symmetric gauge eigenstates. The second
equality follows from Eq.~(\ref{eq:wavefunctions}) and the observation
that $G_{n,m}(0) =0$ for $n \ne m$.

\section{Incompressibility}
\subsection{What is incompressibility?}

In this section of the notes we will argue that incompressibility at
zero temperature in the
absence of disorder is a necessary condition for the
occurrence of the quantum Hall effect. Before beginning, however, it is
useful to state more precisely what we mean by incompressibility.
The compressibility (for a
two-dimensional system at zero temperature) is
defined as the relative area decrease per unit increase in pressure;
\begin{equation}
\kappa \equiv -\frac{1}{V}\frac{\partial V}{\partial P}
\end{equation}
It is usually more convenient to calculate thermodynamic properties as a
function of area rather than as a function of pressure so that the
following expression is often more useful:
\begin{equation}
\kappa^{-1} \equiv -V\frac{\partial P}{\partial V}
 = V\frac{\partial^2E}{\partial V^2}.
\label{eq:compressibility}
\end{equation}
For systems of many particles the energy in the thermodynamic limit is an
extensive quantity and the energy per particle depends on the area and
the particle number only through the particle density, $E = N \epsilon
(n)$ where $n=N/A$ is the areal density. This relationship allow us to
connect the compressibility with the chemical potential;
$\mu = \partial E / \partial N = d(n \epsilon(n)) / d n$. Comparing with
Eq.~(\ref{eq:compressibility}) we find that
\begin{equation}
\kappa^{-1} = n^2\frac{d\mu}{dn}
\end{equation}
When we say that the system is incompressible we mean that
$\kappa=0$. This occurs whenever the chemical potential of the system
increases discontinuously as a function of density.  That is what we
really mean by incompressibility.  I argue below that the quantum Hall
effect can occur whenever the system is incompressible in the
absence of disorder at {\em magnetic field dependent densities}.

Whenever the ground state is incompressible the increase in energy when a
particle is added to the system and the decrease in energy when a
particle is removed from the system differ even in the thermodynamic limit.
It follows that it costs a finite energy to create particle-hole
pairs which are not bound to each other and are therefore able to carry
current.  In this circumstance
we say that the system has a `charge gap'. At this point we
might be tempted to conclude that the chemical potential discontinuity
can be measured by studying the temperature dependence of activated
transport processes in the system. For several reasons this conclusion
would be only partially correct. I will return to this point below.

\subsection{Incompressibility Implies Quantization}
\label{subsec:incomp}

I now present an argument for the following conclusion: {\em
Incompressibility at T=0 in the absence of disorder is a necessary
condition for the occurrence of the quantum Hall effect.
Incompressibility at a magnetic-field dependent density always leads to
gapless excitations localized at the edges of the system. When such an
incompressibility exists in the absence of disorder, the quantum Hall
effect will occur if any gapless excitations which occur in the bulk of
the system are localized on a length scale small compared to the system
size and disorder is not too strong.} Our argument is closely related to
the thermodynamic argument of Widom~\cite{widom} and related arguments due
to Laughlin~\cite{laugauge} and Halperin~\cite{halpstreda}. At least for
non-interacting electrons, similar conclusions follow from treatments
based on linear response theory~\cite{linresp}. I  believe that there are
also close connections between the following argument and the beautiful
topological quantum number picture of the fractional quantum Hall
effect~\cite{avron}, although we don't have the space to pursue these
connections here. There are difficulties with the argument which follows,
and we will touch on some of these difficulties and how they might be
circumvented later. However it is our belief that what follows is the
essence of the quantum Hall effect.

I first consider a large but finite two-dimensional electron gas at zero
temperature, as illustrated in Fig.~[\ref{fig:incomp}],
at a chemical potential which would fall in a `charge gap'in the
thermodynamic limit. I want to consider the change in the equilibrium
local currents, present in the system because of the breaking of
time-reversal-invariance by the magnetic field, when we make an
infinitesimal change in the chemical potential,
$\delta \mu$. I argue that because $\mu$ lies in a gap the change in the
local current density anywhere in the bulk of the system must be zero.
The current density can change, if it does anywhere, only at the edge of
the system. It follows from charge conservation that, if there is a change
in the current flowing along the edge of the system, it must be
constant as we move along the edge. We can relate this change in current
to the change in the orbital magnetization:
\begin{equation}
\delta I = {c \over A} \delta M.
\label{eq:orbmag}
\end{equation}
Eq.~(\ref{eq:orbmag}) is just the equation for the magnetic moment of a
current loop. However,
\begin{equation}
\delta M = {\partial M \over \partial \mu}\vert_B \delta \mu =
{\partial N \over \partial B}\vert_{\mu} \delta \mu.
\label{eq:maxwell}
\end{equation}
The second equality in Eq.~(\ref{eq:maxwell}) follows from a Maxwell
relation. Combining Eq.~(\ref{eq:orbmag}) and Eq.~(\ref{eq:maxwell}) we
obtain the following result for the rate at which the
equilibrium edge current changes with chemical potential when the
chemical potential lies in a charge gap:
\begin{equation}
{\delta I \over \delta \mu} = c {\partial n \over \partial B}|_{\mu}.
\label{eq:streda}
\end{equation}
Notice that whenever the charge gap occurs at a density which depends on
magnetic field, there {\em must} be gapless excitations at the edge of
the system. (These edge states form a chiral one-dimensional electron
system with many interesting properties, especially in the fractional
case~\cite{macdedge,wenedge,fk}. Unfortunately space does not permit us to
review that aspect of the fractional quantum Hall effect theory here.)
{}From a more microscopic point of view, Eq.~(\ref{eq:streda}) arises
because the edge currents are related to the way in which the spectrum
evolves with changes in the the vector potential and hence in the
magnetic field~\cite{halpstreda}. The density at which a charge gap occurs
can depend on magnetic field only if states localized at the edge of the
system cross the Fermi level as a function of magnetic field.

This property of the edge states is expected to persist even if the
chemical potential lies only in a mobility gap and not in a true gap, as
illustrated schematically in Fig.~[\ref{fig:mobility}].
A net current can be carried from source to drain across the system by
changing the local chemical potentials only at the edges and having
different chemical potentials along the two edges connecting source and
drain. Because of localization the two edges and the bulk are
effectively decoupled from each other. Eq.~(\ref{eq:streda}) also
relates the chemical potential difference between the two edges, equal to
$e V_H$,
and the net current carried through the system.
There is no voltage drop along an edge since each edge is in local
equilibrium. Eq.~(\ref{eq:streda}), often called the Streda formula or the
Widom-Streda formula, was first derived from the Kubo formula expression
for the Hall conductivity by Streda and Smrcka~\cite{smrcka} prior to the
experimental discovery of the quantum Hall effect. However, its
robustness over finite ranges of field when localization occurs was
appreciated only after von Klitzing's discovery.

There are difficulties with this argument, most of which are shared with
the elegant Landauer-Buttiker~\cite{buttiker} edge-state picture of
transport in the quantum Hall regime. (The two pictures are essentially
equivalent except that our argument removes unnecessary details specific
to the case of non-interacting electrons. On the other hand the
Landauer-Buttiker picture allows for a natural description of deviations
from the quantum Hall effect in finite-size non-interacting electron
systems.) The principle difficulty with this explanation is that it
would appear to break down when $e V_H$ becomes comparable to the charge
gap or the mobility gap. We know that this is not the case, since
accurately quantized Hall conductances are seen experimentally when $e V_H
$ is hundreds of times larger than $\hbar \omega_c$. Furthermore, the
argument appears at first sight to depend on the assumption that all the
transport current flows at the edges of the system, an assumption which is
certainly not correct in general~\cite{thoulessedge}. These apparent
deficiencies can be explained away by appealing to locality
properties~\cite{niu} of the conductivity when the Fermi level lies in a
region of localized states.

The fact that it is possible to offer various seemingly different
explanations of the quantum Hall effect often creates some confusion. In
closing this section, we wish to emphasize that the different
explanations are really different points of view on the same physics.
For example, the quantization of the Hall conductance is often discussed
in terms of the dependence of properties of the system on boundary
conditions or on Aharanov-Bohm fluxes in different geometries. Because of
localization at edges dependencies on flux are equivalent to dependencies
on magnetic field. For states localized at a particular edge of the system
dependencies on flux are indistinguishable from changes in the magnetic
field strength. A thorough discussion of the connections between the
various approaches taken to explain the quantum Hall effect would be a
good subject for another set of lectures. However that is not the main
subject of these lectures. Instead, we want to emphasize what the
occurrence of the quantum Hall effect tells us about the electronic
properties of a two-dimensional system in which it occurs.  It tells us
that the system has an incompressibility at a magnetic-field-dependent density
in the absence of disorder and that on the Hall plateau all gapless
excitations in the bulk are localized.

\section{Integer Quantum Hall Regime}

 The main focus of these notes will be on the fractional Hall regime. I
first briefly summarize some highlights of the physics of the integer
quantum Hall regime. In the integer quantum Hall regime electron-electron
interactions are weak compared to the electron-disorder interaction.
Experimentally it appears that interactions do not play an essential role
except in the extremely high mobility samples where the fractional quantum
Hall effect occurs, so the integer quantum Hall regime is certainly
realizable experimentally. I have considered the non-interacting problem
in the absence of disorder in detail in a previous section. We found
that the single-particle energies are grouped into macroscopically
degenerate Landau levels. The kinetic eigenenergy for states in the n-th
Landau level is $\hbar \omega_c (n+1/2)$ and the number of states per unit
area in the $n-$th Landau level is $(2 \pi \ell^2)^{-1}$. It is usual and
customary in the quantum Hall regime to measure the charge density in
terms of the charge density of a single Landau level. The Landau level
filling factor is defined by
\begin{equation}
\nu \equiv 2 \pi \ell^2 n.
\label{eq:nu}
\end{equation}
In the absence of disorder discontinuities occur in the chemical
potential of a system of non-interacting electrons at the
magnetic-field-dependent densities corresponding to integer values of
$\nu$.

I have argued in (\ref{subsec:incomp}) that at zero temperature the Hall
conductance will be quantized at
$V_H/I \equiv G_H = j e^2 /h$ and that the dissipative conductance will be
zero over the range of filling factors surrounding
$\nu = j$ where the localization length is finite. Thus experiments on
the integerquantum Hall effect tell us first of all that there is an
incompressibility at a magnetic field dependent density in the absence of
disorder and from the width of the Hall plateau they tell us the range of
filling factors over which states at the Fermi level are localized. The
fact that the Hall conductance is finite tells us that in contrast to the
zero magnetic field situation not all states are localized. Very early
experiments~\cite{mikko} by Paalanen {\em et. al.} showed that in the
integer Hall regime the plateau width approaches the full width of the
Landau level as the temperature approaches zero, implying that extended
states occur at a single critical energy $E_c$ within each Landau level.
Later experiments~\cite{wei,koch} pioneered by Wei and collaborators were
able to extract information about the way in which the localization
length, $\xi$, diverges at the critical energy, $E_c$ within each Landau
level. The experiments are actually performed in samples with
essentially constant density and the position of the Fermi energy within
a Landau level is altered by changing the magnetic field so that what is
studied is
\begin{equation}
\xi (E_F,B) \sim (B - B_c(E_F) )^{-\nu}.
\end{equation}
(Here $\nu$ is {\em not} the filling factor. If you are bothered by this
unfortunate notation you may be consoled by reflecting on the rich
weaving of strands in the tapestry of physical theory which it reflects.
Since $E_c$ is a linear function of $B$ the same critical exponent
applies for the dependence of the localization length on energy at fixed
field.) The analysis of these experiments is based on the notion that
at finite temperature electron-electron or electron-phonon scattering
introduces another relevant length scale
$L_{\phi}$ which is expected to diverge with a power law as the
temperature goes to zero:
\begin{equation}
L_{\phi} \sim T^{-p/2 }
\end{equation}
As indicated by the notation used $L_{\phi}$ is usually thought of as a
phase coherence length of diffusing electrons, although this notion is not
as precise here as it
is at weak magnetic fields where it plays an equally important role in the
theory of quantum corrections to Boltzmann transport
properties~\cite{bal}. In this picture the time between inelastic
scattering events which destroy phase coherence diverges as $T^{-p}$ at
low temperatures.  At finite temperature, dissipation is expected
to occur and the Hall conductance is expected to vary with electron
density or with field only when $\xi(E_F) $ exceeds $L_{\phi}$. The
experiments~\cite{wei} of Wei {\em et. al.} showed that the width of
field regime where dissipation occurs vanishes like $T^{\kappa}$ where
$\kappa \approx 0.42$. The result from experiment, then, is that
\begin{equation}
\kappa = p/ 2 \nu \approx 0.42.
\end{equation}

The principle theoretical problem in the integer Hall regime has been to
understand these experimental results. The fact that extended states
occurred at a single energy with each Landau level was initially
explained~\cite{tsukada,luryi,joynt,iordansky} by considering the limit
of disorder potentials which are smooth on the scale of the magnetic
length. In this limit the eigenstates are localized along equipotential
contours and the localization length is the typical size of a closed
equipotential contour. At low energies the equipotential contours will
surround minima in the random potential which is assumed to have some
finite correlation length. The equipotentials are most easily visualized
by considering the energy of interest as a Fermi energy so that the
equipotential surrounds regions which have been filled by lower energy
electrons. For a given finite system an energy, $E_{c}^{-}$ must
eventually be reached where the filled region first extends from one side
of the sample to the other and the localization length therefore reaches
the sample size. We can perform a similar analysis starting from high
energies where the equipotentials at the Fermi energy surround areas
where the electron states are unoccupied and the localization length is
finite. As the Fermi energy is lowered an energy, $ E_c^{+}$ must
eventually be reached where the unoccupied region extends across the
sample so that the localization length reaches the sample size. It is
clear that $E_c^{+} \ge E_c^{-}$ since the occupied area and the
unoccupied area together comprise the entire area of the sample. One or
the other must extend from one side of the sample to the other. For
$E_c^{-} < E_F < E_c{+}$ there exist both regions of unoccupied states and
regions of occupied states which extend from one side of the sample to
the other. It is clear that this is possible only when the sample size
is comparable to the size of typical equipotential contour. In the limit
of an infinite sample $E_c^{-} = E_c^{+} = E_c $. Using results from
percolation theory Trugman has shown that the size of the typical
equipotential contour $\xi \sim | E - E_c| ^{- \nu}$ with $\nu =4/3$.

The diverging localization length at $E_c$ suggests that the localization
behavior can be viewed as a quantum $(T=0)$ critical phenomenon and that
it should therefore be independent of microscopic details such as the
nature of the disorder potential. On the scale of the diverging
localization length any disorder potential eventually appears to be
rough. This observation has motivated field-theoretical~\cite{pruisken}
treatments of transport at low temperatures near $\nu = 1/2$. The above
results, although they were derived for a smooth disorder potential,
should apply for any disorder potential. Numerical calculations have been
able to test this hypotheses in considerable detail although serious
questions remain. Early work by Ando~\cite{andoloc} convincingly
established that even for zero correlation length disorder potentials the
localization length diverges at a single energy within each Landau level.
Paradoxically the numerical calculations have been most convincing for
short correlation length disorder potentials rather than for smooth
disorder potentials, presumably because the localization length in the
latter case exceeds the system sizes at which calculations are possible.
Later finite-size-scaling analyses~\cite{bodo,chalker} were able to
convincingly establish that the localization length diverges with critical
exponent $\nu =7/3$, in disagreement with the percolation theory results.
It is possible to explain~\cite{percptun} the discrepancy as a correction
due the possibility of tunneling between equipotential contours near
saddle points.

\section{Incompressibility at Fractional Filling Factors}

We are interested in understanding how electron-electron interactions give
rise to incompressibilities at fractional values of the Landau level
filling factor $\nu$. We need to learn to treat interactions between
electrons which are confined to a single Landau level, usually the
lowest ($n=0$) Landau level. It turns out that some important lessons
arise from analyzing the two-body problem in the fractional Hall regime
and so we will start there. All of the discussion in this section is based
on the symmetric gauge discussed previously. I will assume in these
notes that the electrons have been completely spin-polarized by the
magnetic field. The physics of spin or other discrete additional degrees
of freedom in the fractional quantum Hall regime forms a large and
interesting subject~\cite{spin} which we are not able to review here.

\subsection{Two body problem: Haldane pseudopotentials}

We want to find the spectrum of the two-body Hamiltonian. The two-body
wavefunctions could be expanded in terms of
$N=2$ Slater determinants formed from antisymmetric products of two
single-particle states. If the electrons are confined to a finite area
containing $N_{\phi}$ units of magnetic flux the number of two-body
eigenstates is $\sim N_{\phi}^2/2$. In the absence of interactions each
eigenstate has an eigenvalue $\hbar \omega_c$; since the kinetic energy
per particle is trivial constant we usually absorb this constant into the
zero of energy and the Hamiltonian then consists of only the interaction
term which lifts the degeneracy of all two-body states. To solvethe
interacting two-body problem it is useful to transform from the
representation of free individual particle wavefunctions to the
representation of free center-of-mass and relative eigenstates. Using the
complex number representation of two-dimensional coordinates
\begin{eqnarray}
Z &\equiv& (z_1 + z_2 ) /2 \\
z & \equiv & z_1 - z_2.
\end{eqnarray}
Here and below we use the upper case for quantities associated with the
center-of-mass coordinate and the lower case for quantities associated
with the relative coordinate. I define~\cite{velocity} orbit center ladder
operators for the center-of-mass and relative states by
\begin{eqnarray}
b_R^\dagger &=& \frac{b_1^\dagger - b_2^\dagger}{\sqrt{2}} \nonumber \\
b_r^\dagger &=& \frac{b_1^\dagger - b_2^\dagger}{\sqrt{2}}
\label{eq:comrellad}
\end{eqnarray}
so that $[b_r,b_r^\dagger]=[b_R,b_R^\dagger]=1$ and
$[b_r,b_R]=[b_r,b_R^\dagger]=0$. It is easy to verify from
Eq.~(\ref{eq:ladders}) that $b_R$ involves only $Z$ and that its
coordinate space form is identical to that of the individual particle
eigenstates except for the replacements $z_i \rightarrow Z$ and
$\ell \rightarrow \ell_R \equiv \ell/ \sqrt{2}$. Similarly
$b_r$ involves only $z$ and the effective magnetic length for the relative
motion eigenstates is $\ell_r = \sqrt{2} \ell$. The two-body state in the
lowest Landau level with center-of-mass angular momentum $M$ and
relative angular momentum $m$ is:
\begin{equation}
|M\rangle_R |m\rangle_r = { (b_R^\dagger)^M (b_r^\dagger)^m.
\over \sqrt{M! m!} } |0,0\rangle
\label{eq:twobody}
\end{equation}
Comparing Eq.~(\ref{eq:onebody}), Eq.~(\ref{eq:twobody}) and
Eq.~(\ref{eq:comrellad}) it is easy to derive an explicit expression for
the unitary transformation relating the two representations for the
two-body problem. Including the kinetic energy part, the two-body
Hamiltonian may be written as
\begin{eqnarray}
{\cal H} &=& \hbar\omega_c (1 + a_1^\dagger a_1 + a_2^\dagger a_2) +
V(\pol{r}_1 - \pol{r}_2)\nonumber\\
&& \hbar\omega_c (1 + a_R^\dagger a_R + a_r^\dagger a_r) + \sum'_m \vert
m\rangle_r {}_r\langle m | V | m \rangle_r {}_r\langle m\vert\nonumber\\
&& \equiv \sum'_m V_m P_m^{1,2}.
\label{eq:haldane}
\end{eqnarray}
Here $P_m^{1,2}$ projects particles $1$ and $2$ onto a state of relative
angular momentum $m$. (Note that the electrons are restricted to states
of odd relative angular momentum by the antisymmetry requirement on the
two-body wavefunction.) The second form for the right-hand side of
Eq.~(\ref{eq:haldane}) follows from the observation that the interaction
term in the Hamiltonian acts only on the relative-motion
degree-of-freedom. {}From the assumption that the interactions are
isotropic, there is no coupling between states of different total angular
momenta. We see that the Hamiltonian is completely specified by a set of
numbers $V_m$ which are simply the interaction energies for pairs of
particles with relative angular momentum $m$. This parameterization of the
Hamiltonian was first introduced by Haldane~\cite{pseudopots} and these
numbers are known as Haldane pseudopotentials. As we will discuss
further below, most of the physics of interacting electrons in the lowest
Landau level is controlled by the few smallest pseudopotentials and the
fractional quantum Hall effect occurs when we have sufficiently
short-ranged repulsive interactions. Thus the problem of interacting
electrons depends on a finite number of distinct important energy scales
and it is this property which opens up the possibility of chemical
potential jumps due to electron-electron interactions. For short-range
repulsive interactions
$V_m > 0$ and $V_{m'} < V_{m}$ for $m' > m$ since larger relative angular
momenta relative wavefunctions are peaked at larger values of the relative
coordinate where the interaction is weaker. We will find it useful in what
follows to discuss what for many purposes can be considered as the ideal
fractional quantum Hall system, the hard core model, for which $V_1 \ne
0$ and $V_i = 0$ for
$i \ne 1$.

With interactions, then, the two-body spectrum consists of a set of
eigenvalues $E_i = V_{2i-1}$ with degeneracy $g_i = 1 + 2 (N_{\phi} - i)
$ for $i = 1,2,\cdots,N_{\phi}$. The degeneracy $g_i$ is determined by the
requirement that the sum of the relative angular momentum and the
single-particle angular momentum cannot exceed $2 N_{\phi}$. It is
readily verified that the total number of states is
$\sim N_{\phi}^2/2$.

\subsection{Some properties of Haldane pseudopotentials}

I pause here to establish some useful relations involving Haldane
pseudopotentials. Firstly we note that it is possible to derive an
explicit expression for the Haldane pseudopotentials in terms of the
Fourier transform of the electron-electron interaction.
\begin{eqnarray}
V_m \equiv {}_r\langle m\vert V\vert m\rangle_r
 &=& \int \frac{d^2\pol{q}}{(2\pi )^2} V(\pol{q})
 {}_r\langle m\vert e^{i\pol{q}\cdot\pol{r}}\vert m\rangle_r\nonumber\\
&=& \int \frac{d^2\pol{q}}{(2\pi )^2} V(\pol{q}) e^{-q^2} L_m(q^2).
\label{eq:vmvq}
\end{eqnarray}
The final form for the right-hand side of Eq.~(\ref{eq:vmvq}) follows from
Eq.~(\ref{eq:projected}) and the observation that the center of mass and
relative wavefunctions are identical apart from the replacement $\ell
\rightarrow \ell_r = \sqrt{2} \ell$. For the case of an ideal
two-dimensional electron gas $V(q) =2\pi e^2/q$ and from Eq.~(\ref{eq:vmvq})
\begin{equation}
V_m = \frac{e^2}{\ell} \sqrt{\frac{\pi}{4}} \frac{(2 m - 1)!!}{2^mm!}.
\label{eq:vmcoul}
\end{equation}
(This result is actually most easily obtained by doing the integral for
the pseudopotential directly in real space.) Note that for large $m$ the
Coulomb $V_m$ approaches
$(e^2 / \ell) / 2 \sqrt{m}$ as expected since the relative wavefunction
for large $m$ is strongly peaked at a relative separation of $ 2 \ell
\sqrt{m} $. Eq.~(\ref{eq:vmvq}) can be inverted to obtain an expression
for $V(\pol{q})$ in term of the Haldane pseudopotentials by using a
special case of Eq.~(\ref{eq:orthogonality});
\begin{equation}
V(\pol{q}) = 4\pi \sum_m V_m L_m(q^2).
\label{eq:vqvm}
\end{equation}
This equation is often useful for realizing the hard-core model in
numerical calculations.

In these notes we discuss explicitly only the case of the fractional
quantum Hall effect in the lowest Landau level of a semiconductor with an
isotropic band structure and in a magnetic field which is perpendicular
to the two-dimensional layer. It is not, in fact, necessary to be so
restrictive. For example the fractional quantum Hall effect can occur
when the $n-1$ lowest Landau levels are completely filled and we need to
consider the effect of electron-electron interactions among electrons
confined to the $n$-th Landau level. The interaction terms are then
completely specified by the two-body matrix elements in the
$n-th$ Landau level. Looking at a particular Fourier component of the
interaction potential we note that
\begin{eqnarray}
& & \langle n,m'_1: n,m'_2\vert e^{i\pol{q}\cdot (\pol{r}_1 -
\pol{r}_2)}\vert n,m_1, : n,m_2\rangle \nonumber \\
&&= [L_n(q^2/2)]^2 \langle m'_1, m'_2\vert e^{i\pol{q}\cdot (\pol{r}_1 -
\pol{r}_2)}\vert m_1, , m_2\rangle
\end{eqnarray}
where the $n=0$ Landau level index is left implicit on the right hand
side. Thus the fractional quantum Hall effect in the $n$-th Landau level
with interaction potential
$V(\pol{q})$ is equivalent to the fractional quantum Hall effect in the
$n=0$ Landau level with interaction potential $ (L_n(q^2/2))^2
V(\pol{q})$. Anisotropy in the band structure~\cite{haldanebook,zheng},
and the complications associated with the degeneracy~\cite{holes} at the
top of the valence band in $GaAs$ can be accounted for with similar ease.
An interesting example of the quantum Hall effect occurs when the
magnetic field is tilted away from the normal to the two-dimensional
layer. In this case~\cite{nikila}, the system is equivalent to a system
with a perpendicular magnetic field but with a non centro-symmetric
interaction. The concept of a Haldane pseudopotential must then be
generalized to allow matrix elements which are not diagonal in relative
angular momentum and one might hope that the physics would be changed in
important ways. However, there have been numerous experimental
studies~\cite{tiltedfield} of the fractional quantum Hall effect in
tilted magnetic fields and nothing very dramatic occurs as long as the
spin degree of freedom does not play a role.

\subsection{Incompressibility in the Hard-Core Model and Laughlin
Wavefunctions}

Now we are in a position to consider many interacting electrons. The
Hamiltonian is
\begin{equation}
H = \sum_{i<j} \sum_m V_m P_m^{ij}
\label{eq:h}
\end{equation}
The occurrence of chemical potential jumps at fractional Landau level
filling factors (nearly) follows from the following statement. {\em In
the thermodynamic limit the ground state energy of the hard-core model is
zero for $\nu \le 1/3$ and non-zero for $\nu > 1/3$.} The proof of this
statement follows.

Any many-body wavefunction formed completely from electrons in the lowest
Landau level must be a sum of products of the lowest Landau level
one-body wavefunctions. It follows that many-body wavefunctions must take
the form:
\begin{equation}
\Psi [z] = P(z_1, z_2, \cdots , z_N)\; \prod_k \exp{(-\vert z_k\vert^2/4)}
\end{equation}
where $P[z]$ is a polynomial in each of the electronic coordinates. The
hard-core model Hamiltonian is non-negative so that its lowest possible
eigenenergy is zero. Let's assume that
$\Psi [z]$ is an eigenstate of the hard-core model with eigenenergy zero.
Then $\Psi [z]$ must be an eigenstate of $P_1^{ij}$ with eigenvalue zero
for any pair of particles $i$ and $j$. The dependence of $\Psi [z]$ on
$z_i$ and $z_j$ can be reexpressed in terms of the relative and
center-of-mass coordinates for this pair of particles:
\begin{eqnarray}
z_{ij} &\equiv& z_i - z_j\nonumber\\
Z_{ij} &\equiv& (z_i+z_j)/2
\end{eqnarray}
We can expand
\begin{equation}
P[z] = \sum_{m} z_{ij}^m F_m
\end{equation}
where $F_m$ depends on $Z_{ij}$ and on the positions of all the other
particles. Only odd values of $m$ appear in the sum. In order for
$\Psi[z]$ to be annihilated by the hard-core model $F_1$ must be
identically zero. It follows that $z_{ij}^3$ is a factor of $P[z]$ so
that
\begin{equation}
P[z] \propto (z_1 - z_2)^3 (z_1 - z_3)^3\cdots (z_{N-1} - z_N)^3
\label{eq:propto}
\end{equation}
and hence that the maximum power to which each coordinate appears in
$P[z]$ is at least $3 (N-1)$. Since the single-particle orbital with
angular momentum $m$ is localized along a ring which encloses area $2 \pi
\ell^2 (m+1)$, it follows that the area per electron for the state
represented by $\Psi[z]$ for large $N$ is at least $3 (2 \pi \ell)^2$ or
that $\nu^{-1} \ge 3$. This is what we wanted to show. We will have
further occasion to relate the area per electron in the system to the
degree of $P[z]$.  In particular it is useful to note that multiplying
polynomials corresponds to adding areas and hence to adding inverse
filling factors.

The most spatially compact zero-energy eigenstate of the hard-core model
is the one for which the relation sign in Eq.~(\ref{eq:propto}) is an
equal sign. This wavefunction is known as a Laughlin
wavefunction~\cite{laughlin}. More precisely, the total angular momentum
is a good quantum number for interacting electrons in the symmetric gauge
and the total angular momentum is equal to the homogeneous degree of the
polynomial $P[z]$. When the relation in Eq.~(\ref{eq:propto}) is an
equality the degree of $P[z]$ is $M= 3 N (N-1)/2$. Incidentally, the area
per electron for large $N$ can also be deduced from the total homogeneous
degree of a polynomial under the assumption, implicit above in any event,
that the electron density is uniform except near the edge of the system.
The proof of the relation, $A/N = 2 M /N^2$, is left as an exercise. For
$M = 3 N (N-1)/2$ the Laughlin wavefunction is the only zero-energy
eigenstate of the hard-core model. The wavefunction can be generalized to
higher powers of $z_i - z_j$:
\begin{equation}
\Psi_m^L[z] = \prod_{i<j} (z_i - z_j)^m \prod_k \exp{(-\vert
z_k\vert^2/4)}.
\label{eq:lauwf}
\end{equation}
These wavefunctions were first suggested as trial wavefunctions for the
many-electron ground state at $\nu=1/m$ by Laughlin~\cite{laughlin} and
were recognized as exact eigenstates of the hard-core model by Trugman and
Kivelson~\cite{trugkiv}.

For $M=3 N (N-1)/2$, it follows from the above that
$E_0(N) = E_0(N-1)=0$ but $E_0(N+1) > 0$. We can define
$\mu_N^{-} \equiv E_0(N) - E_0(N-1)$ and
$\mu_N^{+} \equiv E_0(N+1) - E_0(N)$ keeping $M=3 N (N-1)/2$.
$\lim_{N \to \infty} \mu_N^{\pm} \equiv \mu^{\pm}
$ gives the chemical potential at Landau level filling factors
infinitesimally larger and infinitesimally smaller than $1/3$.
Evidently, for the hard-core model $\mu^{-}=0$. $\mu^{+}$ is not
known analytically and it may not be immediately evident that it is
finite. However all variational approximations which have been explored
for $E_0(N+1)$ give finite energies even in the limit $N \to \infty$ and it
seems clear that the added electron has a finite probability of being in
a state of relative angular momentum $1$ with one of the existing
electrons. I discuss one such variational wavefunction in the next
section. Numerical exact diagonalization calculations~\cite{gros} have
also convincingly supported this conclusion. There is a chemical potential
jump of $\sim V_1$ at $\nu = 1/3$ and, as we've stated previously, this
leads to the fractional quantum Hall effect.

The hard-core model is not a realistic model for electron-electron
interactions in a two-dimensional electron gas. It is useful to think of
a mapping between Hamiltonians and vectors whose values are lists of
Haldane pseudopotentials. The hard-core model and the model for the real
system are two points in this vector space. I have argued above that
there is a chemical potential jump at $\nu = 1/3$ for the hard-core model
because of the impossibility of avoiding states of relative angular
momentum $1$ between pairs of electrons for $\nu > 1/3$. The chemical
potential jump should be a smooth function of the Haldane
pseudopotentials and there should therefore be a finite volume
surrounding the hard-core model point in this vector space where the jump
remains finite. Experiments showing the fractional quantum Hall effect at
$\nu = 1/3$ can be interpreted as proof that the point in the
pseudopotential vector space describing the physical system lies inside
this volume. Numerical exact diagonalization calculations also
convincingly indicate that this should be the case~\cite{exactdiag}.

\section{Laughlin State Properties}
\subsection{Fractionally Charged Quasiparticles}

Perhaps the most exotic property of incompressible states at fractional
Landau level filling factors, is the fact that they have fractionally
charged excitations. Below we give an argument for the occurrence of
fractionally charged quasiparticles which is closely related to the one
given by Laughlin in his classic paper on the fractional quantum Hall
effect~\cite{laughlin}. I consider a system with an incompressible ground
state at a Landau level filling factor $\nu$; we assume that
$\nu = q/p $ is rational since the case of immediate interest is $\nu =
1/3$ and, in any event, incompressibilities can occur only at rational
filling factors as far as we know. I imagine piercing the
system~\cite{caveatflux} with an infinitely thin solenoid located at a
point we take to be the origin, as illustrated in
Fig.~[\ref{fig:fraccharge}].
The flux through the solenoid is slowly changed from $0$ to
$\Phi_{0} = h c /e$ at which point it is invisible to the electrons and
the solenoid can be removed. The state generated in this way must be an
eigenstate ofthe many electron system. However we can show that this
state has charge $ e q/ p$ added or removed from an area surrounding the
origin. According to the Faraday induction law the time dependent flux
gives rise to an electric field
\begin{equation}
E_\phi = \frac{1}{c}\;\frac{d\phi}{dt}\;\frac{1}{2\pi R}
\end{equation}
directed azimuthally around the solenoid. Since the ground state has an
incompressibility at filling factor $\nu$ it has (on a sufficiently long
length scale) no dissipative conductance and a Hall conductivity
$\sigma_{xy} = \nu e^2 / h$. It follows that, if we look along a ring far
enough from the origin, the azimuthal electric field gives rise to a
radial current,
\begin{equation}
j_r = \nu \frac{e^2}{h}\cdot\frac{1}{c}\;\frac{d\phi}{dt}\cdot\frac{1}{2\pi R}.
\end{equation}
Thus when the solenoid is removed the total change in the charge inside
the ring is
\begin{equation}
Q = 2\pi R\;\int j_r \; dt =
\nu \frac{e^2}{h}\cdot\frac{1}{c}\cdot\frac{hc}{e} = \nu e.
\end{equation}
We have generated an excited state with charge $\nu e$ localized within
the microscopic length scale, $\ell$, of the origin.

For $\nu =1/3$ this procedure gives rise to quasiparticles with charges
$\pm e/3$ created at fixed total electron number. For a fixed number of
electrons the charge comes from the edge of the system, which is removed
to infinity in the thermodynamic limit. In general for $\nu = q/p$ this
procedure generates localized charge $\pm p/q$; the argument cannot
determine the number of fractionally charged quasiparticles created by this
procedure. In general it is expected that the procedure generates
$q$ quasiparticles of charge $1/p$ located near the origin. For fixed area
we can generalize this procedure by creating equal numbers of independent
quasielectrons and quasiholes. The activation energy observed in
transport experiments is the energy to make free quasiparticle-quasihole
pairs, $\Delta$. For $\nu =p/q $ this energy will be $1/q$ as large as the
chemical potential gap, $\Delta_{\mu}$:
\begin{equation}
\Delta = \Delta_{\mu}/q \equiv (\mu^{+}-\mu^{-})/q.
\end{equation}

Explicit trial wavefunctions for states with a single quasielectron or
a single quasihole can be constructed by executing Laughlin's
{\em gedanken} experiment~\cite{macdqp}. The quasielectron and quasihole
states differ qualitatively, as we can see from the following simple
cartoon which contains a considerable degree of truth. Laughlin's {\em
gedanken} experiment for generating quasiparticles has the effect of
increasing the angular momentum of each particle by $1$, in the case of
quasiholes, or in the case of quasielectrons of decreasing the angular
momentum of each particle by $1$. We can estimate the expectation value
for the number operator for each angular momentum in the quasielectron and
quasihole states. In the incompressible ground state $\langle n_k
\rangle_0 = 1/m$ for all $k$. In the quasihole state then $\langle n_k
\rangle_{qh} = \langle n_{k-1} \rangle_0$ for $ k \ge 1$ and $\langle n_0
\rangle_{qh} =0$. The missing charge is all in the $k=0$ state. For
quasielectrons we have to consider the effect of decreasing the angular
momentum by one unit. The $m=0$ state is then raised to the $n=1$ Landau
level and all the part of the wavefunction for which $m=0$ was initially
occupied will be projected away by the {\em gedanken} experiment.  The
wavefunction then needs to be renormalized. The ground state has the
property that whenever the $m=0$ state is occupied the states with
$k=1,\cdots,m-1$ must be unoccupied. Otherwise an electron at the origin
would have a finite probability of having relative angular momentum less
than $m$ with one of the other electrons. In our cartoon for the
quasielectron state we ignore correlations in $\langle n_k' n_k
\rangle_0$ other than those between $k=0$ and $k=1,\cdots,m-1$. Then,
since the states with $k=1,\cdots,m-1$ are unoccupied when $k=0$ is
occupied and they are occupied with overall probability $1/m$ they must be
occupied with probability
$1/(m-1)$ when the $k=0$ state is empty. It follows that in the
quasielectron state $\langle n_k \rangle_{qe} = 1/(m-1) $ for
$k=0,\cdots,m-2$ and is $1/m$ otherwise. The excess charge near the
origin is $1/m$. Note that the quasielectron is localized over a
distance $\sim \sqrt{m} \ell$ whereas the quasihole is localized over a
distance $\sim \ell$. Also note that in the quasielectron state $\langle
n_1 n_0 \rangle \sim (m-1)^{-2}$ independent of the number of electrons.
An electron at the center of the quasielectron has a finite probability of
interacting via a hard-core model so that the chemical potential jump is
finite as claimed previously.

In closing we mention another trial wavefunction for the quasihole state
which was suggested in Laughlin's original~\cite{laughlin} paper:
\begin{equation}
\psi_{qh}^L = \prod_k z_k\;\prod_{i<j} (z_i - z_j)^m\;\prod_\ell \exp{(-\vert
z_\ell\vert^2/4)}
\label{eq:lauqh}
\end{equation}
This trial wavefunction for a quasihole is similar but not identical to
the one discussed above. It has the advantage that some of its properties
can be evaluated using an analogy with classical Coulomb
plasmas which we discuss next.

\subsection{Plasma Analogy}

Some properties of the Laughlin wavefunctions for
incompressible ground states and for fractionally charged quasihole
states can be calculated by exploiting an analogy to classical
two-dimensional plasmas. The analogy is based on interpreting the
coordinate representation quantum mechanical distribution function, {\em
i.e.}, the square of the many-electron wavefunction, as the canonical
ensemble distribution function for a classical system of interacting
particles. For example we write
\begin{equation}
\vert\psi_L\vert^2 = e^{-U_0/k_B T}.
\end{equation}
$U_0$ is defined by this equation. It is convenient (but unnecessary)
to choose $ k_B T = 1/m$ so that
\begin{equation}
U_0 = - 2 m^2 \sum_{i<j} \ln |z_i-z_j |
+ m \sum_k \frac{\vert z_k\vert^2}{2}.
\label{eq:plasmaint}
\end{equation}
The first term on the right hand side describes the interaction between
particles of charge $m$ in a two-dimensional plasma while the second term
describes their interaction with an external electric potential. Noting
that the Laplacian of the second term is a constant we see from the
Poisson equation that this potential arises from a uniform charge
density~\cite{caveatpa},
\begin{equation}
n_B = \frac{1}{2 \pi}.
\end{equation}
Because of the long range of the plasma interaction, overall charge
neutrality is required except within a screening length of the edge of the
electron system. Since the electrons have plasma charge
$m$ this implies that the electron density is
\begin{equation}
n = \frac{1}{2 \pi m}
\end{equation}
and that the density is uniform, as promised previously. The filling
factor $\nu = 1/m$.

A similar argument may be used to deduce the quasihole charge in
Laughlin's explicit wavefunction for the quasihole state. The additional
factor in the wavefunction gives rise to an additional contribution to the
effective potential seen by the plasma particles:
\begin{equation}
U = U_0 + m \sum_k 2\ln{\vert z_k\vert}.
\end{equation}
The additional term corresponds to the interaction with an external unit
charge located at the origin. Because of the long range of the plasma
interaction this charge will be perfectly screened by the charge $m$
plasma particles so that $1/m$ of an electron must be missing from within
a plasma screening length of the origin. We have recovered from this
explicit wavefunction our previous result for the quasihole charge.

\section{Chern-Simons-Landau-Ginzburg Theory}

In recent years, issues connected with the quantum statistics of
interacting particles in two-dimensions have become important themes in
several areas of condensed matter theory, including the theory of the
fractional quantum Hall effect. In this section we review the several
connected ways in which these issues have arisen in fractional quantum
Hall effect theory.

\subsection{Anyons}

The Hamiltonian for a system of identical particles commutes with the
operator whichpermutes the indices of any pair of particles. It
follows that the eigenstates of the many-particle Hamiltonian can be
chosen to be eigenstates of the permutation operator. The usual text-book
argument notes that since the square of the permutation operator for a
particular pair is the identity operator the only possible eigenvalues for
the permutation operator are $+1$ or $-1$ corresponding to Bose and Fermi
statistics respectively. In two-dimensions, however, it is physically
sensible to allow the permutation operator eigenvalue to be any phase factor:
\begin{equation}
\Psi (z_2, z_1, z_3, \cdots , z_N) = e^{\pm i\theta} \Psi (z_1, z_2, z_3,
\cdots , z_N).
\label{eq:anyon}
\end{equation}
Crudely, this is true because in two-dimensions exchange paths for a pair
electrons can be distinguished by a winding number. (The sign choice for
the phase factor in Eq.~(\ref{eq:anyon}) is specified by the sense of the
exchange path.) A continuum of statistics is possible and a given {\em
anyon} system is specified by the statistics angle $\theta$ in
Eq.~(\ref{eq:anyon}). (For an introduction to anyon quantum mechanics see
the article of Canright and Girvin~\cite{cganyon} and work cited therein.)
Since elementary particles live in a three-dimensional world they must
have Fermi or Bose statistics. However, quasiparticles in purely
two-dimensional electronic systems could, at least in principle, have
exotic statistics angles. I'll return to this possibility for the
fractionally charged quasiparticles of the fractionally quantum Hall
effect below.

\subsection{Statistical Transmutation}

One interesting application of the theory of anyons has come from the
observation~\cite{wikz80} that we are free to choose any statistics we
like for two-dimensional particles provided that we appropriately modify
the Hamiltonian. The following transformation changes a system with
statistics angle $\theta$ to one with statistic angle $\theta + \pi \alpha$.
\begin{eqnarray}
\Psi' &=& \prod_{j>k} e^{i\theta_{jk}\alpha}\;
\Psi \nonumber \\
H' &=& \frac{1}{2m} \sum_j \left(\pol{p}_j + \frac{e}{c}\pol{A}_j +
\frac{e}{c}\pol{a}_j\right)^2 + U
\end{eqnarray}
where $\theta_{jk}$ is ${\rm Im} \ln(z_k - z_j)$, $\pol{A}$ is the vector
potential corresponding to the external magnetic field, and the
`statistical vector potential' is
\begin{equation}
\pol{a}_j = \frac{-\hbar c}{e} \alpha \sum_{k\neq j}
\pol{\nabla}_j\theta_{jk}.
\end{equation}
Noting that
\begin{equation}
\oint_k \pol{a}_j\cdot d\pol{\ell} = -\Phi_0\cdot\alpha
\end{equation}
where the integral is around any closed contour surrounding
only particle $k$
we see that the `statistical vector potential' corresponds to a
`statistical magnetic field' for particle $j$
\begin{equation}
\pol{B}_j = \pol{\nabla}\times\pol{a}_j = -\Phi_0\cdot \alpha
\cdot\sum_{k\neq j}
\delta (\pol{r}_j - \pol{r}_k).
\label{eq:statfield}
\end{equation}

This transformation allows us to change the quantum statistics of the
particles at the price of attaching `flux tubes' to each particle as
illustrated schematically in Fig.~[\ref{fig:fluxattach}]
and has become known as statistical transmutation. If each flux tube
contains a single magnetic flux quantum and we attach an odd number of
flux tubes we change fermions to bosons. The resulting particles are
often referred to as composite bosons because of the attached
`flux-tubes'. We will have occasion later to discuss transformations in
which even numbers of flux-tubes are attached to each particle, a
transformation which does not alter particle statistics. In that case the
transformed particles are often referred to as `composite fermions'.
Statistical transmutation is the two-dimensional analog of the familiar
Jordan-Wigner transformation in one-dimensional identical particle systems.

\subsection{Quasiparticle Statistics}

It has been argued~\cite{halpqpstat,arovasqpstat} that the quasiparticles
of the $\nu =1/m$ incompressible state have fractional statistics with
statistics angle $\pm 1/m$. This assertion is motivated by trial
wavefunctions for many quasiparticle states. Consider Laughlin's trial
wavefunction for a state with a single quasihole located at $z_0$:
\begin{equation}
\Psi_{z_0}[z] = \prod_i (z_i - z_0) \Phi [z] = \vert\Psi_{z_0}[z]\vert
e^{i\varphi [z]}.
\label{eq:qpz0}
\end{equation}
We can easily calculate the total phase change in this wavefunction when
the quasiparticle is moved around a closed loop keeping all the particles
fixed.
\begin{equation}
\oint dz_0 \frac{d\varphi [z]}{dz_0} = -i\oint
\frac{\psi'_{z_0}[z]}{\psi_{z_0}[z]} dz_0 = 2\pi N_0.
\end{equation}
where $N_0$ is the number of electrons, and hence the number of zeroes of
$\Psi_{z_0}[z]$ (considered as a function of $z_0$) inside the loop. If a
quasiparticle located well in the interior of the loop is introduced in
$\Phi [z]$ in Eq.~(\ref{eq:qpz0}) the additional phase change, averaging
over all electron coordinates, is $2 \pi/m$ since the quasiparticle charge
is $1/m$. Arovas {\em et. al.} have noted~\cite{arovasqpstat} that the
same phase appears in the wavefunction of a pair of fractional statistics
particles where it can be considered as a Berry phase associated with a
statistical vector potential and have concluded that the quasiparticles
are anyons having statistics angle $ \pi /m$.
There is some evidence from finite size exact
diagonalization calculations in favor of fractional quasiparticle
statistics~\cite{beran}. There have not been any suggestions of feasible
experimental measurements which could establish the quasiparticle
statistics. It may seem surprising that the statistics of the
quasiparticles does not give rise to qualitative physical effects.
I'll explain why this is not the case later in this section.

\subsection{Digression: Lowest Landau Level Density Matrices}

I pause here to calculate the ground state one-body density matrix in a
coordinate representation for many-body states in the lowest Landau
level. It is useful to start by considering its diagonal elements the
density, $n (\pol{r})$.
\begin{eqnarray}
n(\pol{r}) &=& \sum_{m',m} \bar{\varphi}_{m'}(z)\varphi_m(z) \langle
c_{m'}^\dagger c_m\rangle_0\nonumber\\
&=& \frac{1}{2\pi} \sum_{m',m} \frac{r^{m+m'} e^{i\theta
(m-m')}}{\sqrt{2^{m+m'} m!m'!}} \langle c_{m'}^\dagger  c_m\rangle_0
e^{-r^2/2}.
\label{eq:2qdens}
\end{eqnarray}
The second form of Eq.~(\ref{eq:2qdens}) follows by explicitly
substituting the symmetric gauge lowest Landau level wavefunctions. We see
that if the ground state represents an isotropic fluid
\begin{equation}
\langle c_{m'}^\dagger c_m\rangle_0\propto \delta_{m',m}
\end{equation}
so that
\begin{equation}
n(r) = \frac{1}{2\pi} \sum_m \frac{!}{m!}\left(\frac{r^2}{2}\right)^m
e^{-r^2/2} \langle n_m\rangle_0.
\end{equation}
If we further assume that the ground state is a constant density fluid at
a Landau level filling factor $\nu$, the density is $\nu / 2 \pi$ and
\begin{equation}
\langle n_m\rangle_0 = \nu
\end{equation}
independent of $m$. The second quantized form for the one-body density
matrix differs only in the position at which $\bar \varphi_m$ is evaluated:
\begin{equation}
\rho (z, \bar{z'}) = \sum_{m',m} \bar{\varphi}_{m'}(z') \varphi_m(z) \langle
c_{m'}^\dagger c_m\rangle_0.
\end{equation}
Using the above result for $\langle c_{m'}^\dagger c_m \rangle_0$ we find
\begin{eqnarray}
n(z,\bar{z'}) &=& \frac{\nu}{2\pi} \sum_m
\frac{1}{m!}\left(\frac{z\bar{z'}}{2}\right)^m e^{-\vert z\vert^2/4}
e^{-\vert z'\vert^2/4}\nonumber\\
&=& \frac{\nu}{2\pi} \exp{\left( -\frac{\vert z - z'\vert^2}{4}\right)}
\exp{\biggl((z\bar{z'} - z'\bar{z})/4\biggr)}.
\label{eq:isodenm}
\end{eqnarray}
The last factor in Eq.~(\ref{eq:isodenm}) is a phase factor. The one-body
density matrix has a Gaussian fall off and is completely specified by
$\nu$ for any isotropic fluid.

\subsection{Boson Off-Diagonal Long-Range Order}

Long-range-order in the one-body density matrix, or in two-dimensions
quasi-long-range order, is the usual criterion for superfluidity. (For
superconductors long range order occurs in the density-matrix of Cooper
pairs.) I have just shown that in the fractional Hall regime long range
order {\em does not} occur in the electron one-body density matrix. There
is, however, an important connection between superfluidity and the
fractional quantum Hall effect which we will now explain. For $\nu = 1/m$
the connection is based on the the statistical transmutation
transformation
\begin{equation}
\Psi^{'}[z] = \prod_{j<k} \exp^{-i m \theta_{jk}} \Psi [z]
\end{equation}
which changes fermions into bosons since $m$ is odd. We want to consider
the one-body density matrix in this boson representation. In first
quantized form the density matrix is:
\begin{equation}
n (z, z') = N\int d^2 z_2\cdots\int d^2z_N \psi (z, z_2, \cdots , z_N)
\bar{\psi} (z', z_2, \cdots z_N).
\end{equation}
For Laughlin's wavefunction the effect of the statistical transmutation is
to replace the wavefunction by its absolute value. The explicit form of
the transformed density matrix is
\begin{eqnarray}
n' (z, z') &=& N\int d^2 z_2\cdots\int d^2z_N \psi' (z: [z]) \bar{\psi'}
(z'; [z])\nonumber\\
&=& N\int d^2z_2\cdots\int d^2z_N \prod_{1<k} \vert z - z_k\vert^m \vert
z' - z_k\vert^m e^{-\vert z\vert^2/4}\nonumber \\
&& e^{-\vert z'\vert^2/4} \prod_{1 < \ell <m} \vert z_\ell - z_m\vert^{2m}
\prod_{1<j} e^{-\vert z_j\vert^2/2}.
\label{eq:bosdm}
\end{eqnarray}
$n' (z,z')$ clearly drops off much more slowly with $|z - z'|$ than
$n (z,z')$ since the integrand in Eq.~(\ref{eq:bosdm}) is real and
positive definite. It is possible to show~\cite{girv87} using a plasma
analogy argument that
\begin{equation}
n' (z,z') \sim |z - z'| ^{-m/2}.
\end{equation}
(I leave this proof as an exercise for the reader.) Thus in the
appropriate boson representation the one-body density matrix of the
incompressible ground state has the quasi-long-range order which is
associated with a two-dimensional superfluid in the absence of a magnetic
field.

This observation has given rise to a useful phenomenology for the
fractional quantum Hall effect known as Chern-Simons-Landau-Ginzburg
theory~\cite{screview}. For $\nu = 1/m$, $m$ quanta of magnetic flux
pass through the system for each electron. We can choose a statistical
transmutation transformation which attaches
$m$ flux quanta to each electron whose orientation is opposite to that of
the physical magnetic field. (In a Lagrangian formulation these attached
flux quanta lead to a Chern-Simons term in the action.) This statistical
magnetic field fluctuates in a complicated way in concert with quantum
fluctuations in the electronic motion. If the fluctuating magnetic field
is treated exactly, something we are not able to do at present, this
description of bosons interacting with each other and with a fluctuating
magnetic field is entirely equivalent to the direct fermionic description
of the interacting electron system. The simplest approximation is to
replace the fluctuating magnetic field by its average, yielding a system
of bosons with repulsive interactions in zero magnetic field. Bosons
with repulsive interactions are superfluids and therefore share the
off-diagonal-long-range-order of the Laughlin state transformed to the
boson representation. It has been argued~\cite{girv87} that this
long-range-order property occurs in the ground state if and only if the
fractional quantum Hall effect occurs, and therefore that the fractional
quantum Hall effect is equivalent to superfluidity in the appropriate
boson representation. Given that the boson system is a superfluid and that
fluctuations in the statistical magnetic field can be treated at a
random-phase-approximation level, all essential physical properties
associated with the fractional quantum Hall effect can be explained on a
phenomenological level~\cite{cslg}. It is important to realize that the
validity of the random-phase-approximation treatment of the fluctuating
magnetic field cannot be justified on theoretical grounds. It is
evidently a poor approximation in the case of non-interacting electrons
since nothing like the macroscopic ground state degeneracy of that limit
can be recovered. Its apparent validity is evidently due to the same
correlations which give rise to the fractional quantum Hall effect.

The polynomial part of the Laughlin state at $\nu = 1/m$ is of degree $m
(N-1)$ in each of its coordinates, and hence has $m (N-1)$ zeroes as a
function of any of its coordinates. The Laughlin state has the special
property that $m$ zeroes are placed at the position of each other particle
and they follow these particles as they move about. This is the property
of the Laughlin states which makes them have a low energy for short-range
repulsive interactions. It is also this property of the Laughlin
wavefunctions which makes their boson transformed counterparts purely real
and leads to off-diagonal-long-range-order and the order parameter of the
Chern-Simons-Landau-Ginzburg theory. A closely related order parameter has
been constructed by Read~\cite{read} and numerical calculations by Rezayi
and Haldane~\cite{edorder} have verified that this order exists if and
only if the electron-electron interaction is such that the fractional
quantum Hall effect occurs.

\section{Collective Modes of Incompressible States}

\subsection{Collective Modes at Zero Magnetic Field}

In this section we will discuss the collective modes and the
density-density response function at zero temperature for the case where
the ground state is incompressible. Our approach is similar to sum rule
approaches which are often useful for discussing collective modes and
response functions in interacting particle systems. A famous example is
Feynman's theory~\cite{feynman} of the collective excitations and response
functions of ${}^3{\rm He}$. To orient ourselves to the problem at hand we
consider the application of a similar approach to the case of interacting
two-dimensional electrons in the absence of a magnetic field. The
central quantity in this approach is the dynamic structure factor of the
electron system which is defined by
\begin{equation}
s(q,\epsilon) \equiv N^{-1} \sum_n | \langle \Psi_n | \rho_{\pol{q}} | \Psi_0
\rangle |^2 \delta (\epsilon - (E_n - E_0)).
\label{eq:dsf}
\end{equation}
Here $\rho_{\pol{q}}$ is the Fourier transform of the one-body density
operator;
\begin{equation}
\rho_{\pol{q}} \equiv \int d \pol{r}
 \exp (- i \pol{q} \cdot \pol {r} ) \sum_{i=1}^N \delta ( \pol{r} - \pol{r}_i )
\end{equation}
(We use the notation $n(\pol{q}) = \langle \Psi_0 | \rho_{\pol{q}} |
\Psi_0 \rangle$ to distinguish the density operator from its ground state
expectation value.) Three different moments of the dynamic structure
factor have important physical significance:
\begin{eqnarray}
f(q) &\equiv& \int d \epsilon \epsilon s(q,\epsilon) \\
s(q) &\equiv& \int d \epsilon s(q,\epsilon) \\
\chi(q) & = & 2 n \int d \epsilon \frac{s(q,\epsilon)}{\epsilon}.
\label{eq:moments}
\end{eqnarray}
Both $f(q)$ and $s(q)$ can be expressed in terms of ground state
expectation values of appropriate operators:
\begin{equation}
f(q) = N^{-1} \langle \Psi_0 | \rho_{-\pol{q}} [ H, \rho_{\pol{q}}] |
 \Psi_0 \rangle = \frac{\hbar^2 q^2}{2 m}
\label{eq:fsumrule}
\end{equation}
and
\begin{equation}
s(q) = N^{-1} \langle \Psi_0 | \rho_{- \pol{q}} \rho_{\pol{q}} | \Psi_0
\rangle.
\label{eq:statsf}
\end{equation}
The second equality in Eq.~(\ref{eq:fsumrule}) is known as the f-sum
rule. The first moment of the dynamic structure factor turns out to be
proportional to $q^2$ and completely independent of the electron electron
interaction. It can be shown~\cite{pinesnozieres} that $\chi (q)$ is the
static density-density response function.

Because of the long range interactions in interacting electron systems,
some general statements can be made about the behavior of $\chi(q)$ at
small $q$. This is most easily seen by considering the energy of the
electronic system as a functional of the Fourier components of the ground
state electronic density. This assumption can be formally justified by
appealing to the Hohenberg-Kohn theorems of density-functional
theory~\cite{hkthms}. If we consider the charge density induced in the
uniform system by a weak external potential $V^{ext}(\pol{q})$ we find
that
\begin{equation}
\chi (q) \equiv - \frac{n(\pol{q})}{V^{ext}(\pol{q})} =
 \frac{1}{A} \big( \frac{\delta^2 E^{int}}
{\delta n(\pol{q}) \delta n(\pol{-q})} \big)^{-1}.
\label{suscep}
\end{equation}
The response function, $\chi (q)$, is inversely proportional to the
stiffness of the internal energy of the electron system when density
modulation is introduced at wavevector $\pol{q}$. For systems with long
range interactions it is customary to separate the contribution to the
stiffness from the Hartree energy of the charge density so that for a
two-dimensional electron system
\begin{equation}
 \chi^{-1} (q) = \frac{ 2 \pi e^2 }{ q} + \Pi^{-1}(q).
\label{eq:inversuscep}
\end{equation}
We may take the static polarization function, $\Pi (q)$ to be defined by
Eq.~(\ref{eq:inversuscep}). It can be demonstrated from the above
discussion that
\begin{equation}
\lim_{q \to 0} \Pi^{-1} (q) = \frac{d \mu}{ d n}
\end{equation}
where the energy of the system is calculated with a neutralizing
background at every density. (This connection between $s(q,\epsilon)$ and
$\chi (q)$ is often referred to as the compressibility sum rule. For
short range interactions the connection is more direct since the
Hartree-interaction need not be separated.) As long as $\lim_{q \to 0}
\Pi (q)$ is finite, a result which is believed to apply whenever the
two-dimensional electron system has a fluid ground state, we can conclude
that
\begin{equation}
\lim_{q \to 0} \chi (q) = \frac{q}{2 \pi e^2}.
\end{equation}
Independent of any assumptions about the long wavelength behavior of
$\Pi (q)$ it follows from the above arguments that $\chi (q)$ must vanish
at least like $q$ for $q$ going to zero. A two-dimensional electron
system cannot change its density on long length scales because of the
infinite energy cost of violating overall charge neutrality. This
behavior is sometimes referred to as incompressibility since $d \mu / d n$
is infinite unless the neutralizing positive background charge density is
changed along with the electron density. We see below that the
incompressible states associated with the fractional quantum hall effect
are `much more' incompressible since $\chi (q)$ vanishes like $q^4$ at
small $q$.

We can use the above results to estimate the elementary excitation
energies and the response functions of the  two-dimensional electron
system. The estimates are based on the assumption that there is a single
excited state which contributes to $s(q,\epsilon)$ at each $q$; this
approximation is known as the single-mode approximation. The assumption is
well justified in the present case for $q \to 0$ since in this limit
$\rho \propto \pol{q} \cdot (\sum_i r_i) $, (up to a constant) which
involves only the center of mass coordinate of the electrons. The
center-of-mass motion is completely decoupled from the complicated
relative motion of the particles of the gas and its excited states are
labeled by momentum. For larger $q$ we know that many individual
particle-hole pair excitations are possible at a given $q$ and the
single-mode-approximation should be regarded critically. Given the single
mode approximation we can compare our exact results for $\chi (q)$ and $
f(q)$ to obtain an estimate of the energy of the collective excited state:
\begin{equation}
\epsilon_{pl}(q) = \big( \frac{2 \pi n e^2 q \hbar^2 }{m} \big)^{1/2}.
\label{eq:engplas}
\end{equation}
This is, of course, the long-wavelength expression for the plasmon
collective excitation of the zero-field two-dimensional electron gas.
Given Eq.~(\ref{eq:engplas}) it follows that for
$q \to 0$
\begin{equation}
S ( q, \epsilon) = \frac{ \hbar ^2 q^2} { 2 m \epsilon_{pl}(q)}
\delta (\epsilon - \epsilon_{pl}(q))
\end{equation}
so that $S(q) \sim q^{3/2} $ at small $q$ compared to the familiar result
that for non-interacting electrons $S(q) \sim q$.

\subsection{Correlation Function Moments in the Fractional Hall Regime}

I now want to apply a similar analysis in the fractional Hall regime.
We'll find that our conclusions depend on some general properties of
uniform fluid state correlation functions which we now establish.
Consider the second quantization expression for the two-point distribution
function
\begin{eqnarray}
& & n^{(2)}(\pol{r}_1, \pol{r}_2) \nonumber \\
&=& \sum_{{m_1 , m_2}\atop {m'_1, m'_2}}
\bar{\varphi}_{m'_1} (\pol{r}_1) \varphi_{m_1 } (\pol{r}_1)
\bar{\varphi}_{m'_2} (\pol{r}_2) \varphi_{m_2} (\pol{r}_2) \langle
c_{m'_1}^\dagger c_{m'_2}^\dagger c_{m_2} c_{m_1} \rangle_0
\end{eqnarray}
Assuming that the ground state is a uniform isotropic fluid,
$n^{(2)}(\pol{r}_1, \pol{r}_2) = n^2 g(|\pol{r}_1 - \pol{r}_2 |) $ where
$g(r)$ is the usual dimensionless pair distribution function normalized
so that it goes to one at large $r$. We can evaluate $g(r)$ by choosing
$\pol{r}_1 = 0$ and
$\pol{r}_2 = r \hat x$. We take advantage of the frequently convenient
fact that only the $m=0$ orbital is non-zero at the origin to find,
\begin{equation}
g(r) = n^{-2} \sum_{m\neq 0}
\left(\frac{r^2}{2}\right)^m \frac{1}{m!} e^{-r^2/2}
\langle\hat{n}_m\hat{n}_0\rangle_0.
\end{equation}
In the following we define $x \equiv r^2 /2$. It it possible to derive
general expressions for two spatial moments of the dimensionless pair
correlation function
\begin{eqnarray}
h(r) &\equiv& g(r) - 1 \nonumber\\
&=& {\nu}^{-2} \sum_{m=0}^\infty\frac{x^m}{m!} \, e^{-x}
\biggl((1 - \delta_{m,0})\langle\hat{n}_m\hat{n}_0\rangle -
\langle\hat{n}_m\rangle\langle\hat{n}_0\rangle\biggr).
\end{eqnarray}
The zeroth moment of $h(r)$ is
\begin{equation}
n\int d^2\pol{r} h(r) = \nu^{-1}\biggl[\langle\hat{N}\hat{n}_0\rangle -
\langle\hat{n}_0\hat{n}_0\rangle - \langle\hat{N}\rangle
\langle\hat{n}_0\rangle\biggr] = -1,
\end{equation}
where the the second equality follows from the fact that the total number
operator is a good quantum number. The first moment of $h(r)$ is
\begin{eqnarray}
n\int d^2\pol{r} \frac{r^2}{2} h(r) &=& -1 + \nu^{-1}
\biggl[\langle\hat{M}\hat{n}_0\rangle - \langle\hat{M}\rangle
\langle\hat{n}_0\rangle\biggr]\nonumber\\
&=& -1
\end{eqnarray}
where $\hat{M} = \sum_m m\hat{n}_m$ and the second equality this time
follows from the fact that the total angular momentum is a good quantum
number. These two sum rules for the pair correlation function apply to
Laughlin's wavefunction. In that special case they can be derived from
the plasma analogy using perfect screening properties of the plasma but we
see here that they have a more general validity. The fact that these sum
rules apply to any many-electron wavefunction in the lowest Landau level
which represents an isotropic fluid, suggests that there is a deep
connection between the suppression of long wavelength density fluctuations
in two-dimensional plasmas by long-range interactions and the suppression
of density fluctuations in the fractional quantum Hall regime by
projection onto a single Landau level.

\subsection{Projected Static Structure Factor}

The pair correlation function and the static structure factor discussed
above are closely related. To see this recall that
\begin{equation}
s(k) = \frac{1}{N}
\langle\rho_{-\pol{k}} \rho_{\pol{k}}\rangle_0 =
 \left\langle\frac{1}{N} \sum_{i, j} e^{i\pol{k}\cdot
(\pol{r}_i - \pol{r}_j)}\right\rangle_0.
\label{eq:skcf}
\end{equation}
The $i=j$ terms in Eq.~(\ref{eq:skcf}) sum to $1$ and the $i \ne j$ terms
are proportional to a Fourier transform of the two-point distribution
function:
\begin{eqnarray}
s(k) &=& 1 + \frac{1}{N} \int d\pol{r} \int d\pol{r'} e^{i\pol{k}\cdot
(\pol{r} - \pol{r'})} n^{(2)} (\pol{r},
\pol{r'})\nonumber\\
&=& 1 + N\delta_{\pol{k},0} + n\int d\pol{r} e^{i\pol{k}\cdot\pol{r}} h(r)
\equiv 1 + N \delta_{\pol{k},0} + h(k)
\label{eq:skcf2}
\end{eqnarray}
In Eq.~(\ref{eq:skcf2}) we have used the the first quantization expression
for the two-point distribution function:
\begin{equation}
n^{(2)} (\pol{r}, \pol{r'}) \equiv \sum_{i\neq j} \biggl\langle\delta
(\pol{r} - \pol{r}_j) \delta (\pol{r'} - \pol{r}_j)\biggr\rangle_0
\end{equation}
and adopted the conventional definition for $h(k)$, the Fourier transform
of the pair correlation function. The moments derived above for $h(r)$
imply the following general result for the long-wavelength behavior of
$h(k)$.
\begin{eqnarray}
h(k) &=& n\int d\pol{r} h(r) + \frac{k^2}{2} \left( -n\int dr \frac{r^2}{2}
h(r)\right) + {\cal O} ( k^4)\nonumber\\
&=& -1 + \frac{k^2}{2} + \cdots.
\label{eq:hklong}
\end{eqnarray}

In the fractional Hall regime, we are interested primarily in the low
energy elementary excitations of the system which do not involve the
promotion of electrons to higher Landau levels. It is explicit in
Eq.~(\ref{eq:pwme}) that $\rho_{\pol{k}}$ maps states in the lowest Landau
level partly to higher Landau levels. To generate the excited states of
interest it is necessary to project the density operator onto the lowest
Landau level:
\begin{equation}
\bar{\rho_{\pol{k}}} \equiv \sum_i {}_i\langle 0\vert
e^{-i\pol{k}\cdot\pol{r}_i}\vert 0\rangle_i = \sum_i B_i(k)
\end{equation}
where $B_i(k)$ was expressed in terms of the orbit center ladder
operators for particle $i$ in Eq.~(\ref{eq:pwproj}). Note that
\begin{equation}
B_i(k_1) B_i(k_2) = e^{k_1\bar{k}_2/2} B_i(k_1 + k_2)
\end{equation}
so that $B_i (-k) B_i (k) = \exp ( - |k|^2/2 ) $. The projected static
structure factor for $\pol{k} \ne 0$ obeys
\begin{eqnarray}
\bar{s}(k) &\equiv & \frac{1}{N} \langle\bar{\rho}_{-\pol{k}}
\bar{\rho}_{\pol{k}} \rangle = \frac{1}{N} \sum_{i\neq j} \langle
e^{i\pol{k}\cdot\pol{r}_i} e^{-i\pol{k}\cdot\pol{r}_j}\rangle_0 +
\frac{1}{N}\sum_i \langle B_i(-k) B_i(k)\rangle\nonumber\\
&=& s(k) - 1 + e^{-\vert k\vert^2/2} = h(k) + e^{-\vert k\vert^2/2}
\label{eq:sbarhk}
\end{eqnarray}
To obtain this result we've noted that the ground state is entirely in
the lowest Landau level so that the projection of
$\exp ( - i \pol{k} \cdot \pol{r} ) $ is necessary only when the two
particle indices are identical. Combining Eq.~(\ref{eq:hklong}) with
Eq.~(\ref{eq:sbarhk}) we obtain the remarkable result that
\begin{equation}
\bar{s}(k) = \left( -1 + \frac{\vert k\vert^2}{2} + {\cal O} (\vert
k\vert^4)\right) + \left( 1 - \frac{\vert k\vert^2}{2} + {\cal O} (\vert
k\vert^4)\right).
\end{equation}
Inserting the definition of the projected static structure factor we can
conclude that there are no dipole matrix elements within the lowest
Landau level:
\begin{equation}
\sum_m \left\vert\langle\psi_m\vert
e^{-i\pol{k}\cdot\pol{r}}\vert\psi_0\rangle\right\vert^2 = {\cal O}
 (\vert k\vert^4)
\end{equation}
By dipole matrix elements we mean those that would arise at the lowest
order in the expansion of $\exp ( -i \pol{k} \cdot \pol{r} )$ (if this were
valid) and give rise to contributions to $\bar s (k) \sim k^2$. (For
finite systems dipole matrix elements do occur within the lowest Landau
level but only for excitations that are localized at the edge of the
system.) Note that since the real part of the conductivity is related to
$s(q,\omega)$ by the continuity equation, it follows that $\sigma_{xx}
(\omega) \equiv 0$ within the lowest Landau level if no disorder potential
is present.

\subsection{Magnetorotons}

We are now in a position to discuss the intra-Landau-level collective
modes of the fractional quantum Hall effect using the single-mode
approximation. The calculation we outline below was first performed by
Girvin {\em et al.}~\cite{girvsma} The calculation is based on a version
of the single-mode-approximation projected onto the lowest Landau level.
These modes have become known as magnetorotons because their existence was
first suggested based on a strong-magnetic-field generalization of
Feynman's theory of the $He^3$ phonon-roton excitation spectrum. We assume
that $\vert\psi_{\pol{k}} \rangle =
\bar{\rho}_{\pol{k}}\vert\psi_0\rangle$ is an approximate eigenstate of
the many-electron Hamiltonian and attempt to evaluate its energy and the
matrix element of the (projected) density operator between this state and
the ground state so that we can estimate response functions. The energy is
\begin{equation}
E(\pol{k}) = \frac{\langle\psi_{\pol{k}}\vert {\cal
H}\vert\psi_{\pol{k}}\rangle}{\langle\psi_{\pol{k}}\vert\psi_{\pol{k}}\rangle}
= E_0 + \frac{\langle\psi_0\vert\bar{\rho}_{-\pol{k}} [{\cal H},
\bar{\rho}_{\pol{k}}]\vert\psi_0\rangle}{\langle\psi_0\vert\bar{\rho}_{-\pol{k}}
\bar{\rho}_{\pol{k}}\vert\psi_0\rangle} \equiv
E_0 + \frac{\bar{f}(\pol{k})}{\bar{s}(\pol{k})}.
\label{eq:mreng}
\end{equation}
In Eq.~(\ref{eq:mreng}) $\bar f(k)$ is the projected version of the
usual f-sum rule.
\begin{equation}
\bar{f}(k) \equiv \frac{1}{N} \sum_m
\vert\langle\psi_m\vert\bar{\rho}_{\pol{k}}\vert\psi_0\rangle\vert^2 (E_m -
E_0) = \bar{f}(-k).
\end{equation}
The second equality follows from inversion symmetry in the isotropic
fluid and it allows us to write
\begin{equation}
\bar{f}(k) = \frac{1}{2N} \langle\psi_0\vert [\bar{\rho}_{-k}, [{\cal H},
\rho_k]]\vert\psi_0\rangle.
\end{equation}
The Hamiltonian consists of the interaction term alone which can be
written in the form
\begin{equation}
{\cal H} = \frac{1}{2} \int \frac{d^2\pol{q}}{(2\pi )^2} v(q) \bar{\rho}_{-q}
\bar{\rho}_q + \hbox{ constant}.
\end{equation}
It is amusing to note that for the usual unprojected f-sum rule the entire
contribution comes from the kinetic energy term in the Hamiltonian. Here
however there is a contribution from the interaction term because the
projected density operators do not commute,
\begin{equation}
 [\bar \rho_{k_1}, \bar \rho_{k_2}] = \sum_i [B_i(k_1), B_i(k_2)]
= (e^{k_1 \bar k_2/2} - e^{k_2 \bar k_1/2} ) \bar \rho_{k_1+k_2}
\label{eq:pdocom}
\end{equation}
After a little patient work using Eq.~(\ref{eq:pdocom}), the expression
for $\bar f (k)$ can be expressed in terms of the projected static
structure factor.
\begin{equation}
\bar{f}(k) = \int \frac{d^2\pol{q}}{(2\pi )^2} v(q) (1 - \cos{(\hat{z}\cdot
(\pol{q}\times\pol{k}))}) (\tilde{\bar{s}}(k+q) - \tilde{\bar{s}}(q))
e^{-\vert q\vert^2/2}
\label{eq:mrexplicit}
\end{equation}
where
\begin{equation}
\tilde{\bar{s}}(q) \equiv e^{\vert q\vert^2/2} \bar{s}(q).
\end{equation}

For an isotropic fluid it follows from Eq.~(\ref{eq:mrexplicit}) that
$\bar f(k) \sim k^4$ at small $k$. Since $\bar s(k) \sim k^4$ the
single-mode approximation gives collective excitation energies which
remain finite in the long wavelength limit, unlike the two-dimensional
plasmon excitations at zero magnetic field. For incompressible ground
states there is therefore a gap for both neutral and charged excitations. The
single-mode-approximation magnetoroton excitation energies calculated by
Girvin {\em et. al.} are shown in Fig.~[\ref{fig:mr}]. The excitation
energies are expressed in terms of the natural energy unit in the strong
magnetic field limit for $1/r$ interactions, $e^2 / \ell$.
Because of the excitation gap, $\chi (k)$ also vanishes like $k^4$, much
faster than the linear in $k$ behavior required by long-range interactions
at zero magnetic field. (Actually there is a quadratic contribution to
$\chi (k)$ which comes from higher Landau level excitations but it is
proportional to $B^{-1}$ in the strong magnetic field limit.)

Because $\bar s (k) \sim k^4$ at long wavelengths the magnetoroton is not
purely a center-of-mass motion excitation. We cannot offer a compelling
argument in favor of the accuracy of the single-mode approximation even
in the long wavelength limit, however we do get some indication that is is
likely to be accurate for incompressible ground states by comparing with
the $\nu = 1$ case. The following discussion follows a line of thought
first advanced by Kallin and Halperin~\cite{kalhalp}. At $\nu =1$ the
ground state in the strong magnetic field limit is the single Slater
determinant which has all the single-particle orbitals in the lowest
Landau level occupied. The ground state is incompressible and, for
non-interacting electrons, the lowest energy neutral excitations are the
$N_{\phi}^2$ particle-hole pairs all of which have excitation
energy $\hbar \omega_c$. This degeneracy is lifted by electron-electron
interactions but the excitation energies can still be evaluated exactly
since they are labeled by wavevector and there is precisely one state at
each wavevector. The excited state is
\begin{equation}
|\Psi_{\pol{k}} \rangle \equiv
\sum_i |n=1\rangle_i \langle n=0 | B_i (k) |\Psi_0\rangle.
\label{eq:mgplas}
\end{equation}
The energy of this state can easily be evaluated exactly; the
calculation is similar to that used above to calculate magnetoroton
energies~\cite{macdmgplas}. For large $|k|$ it turns out that
\begin{equation}
E(k) = \Delta \mu - e^2 / k \ell^2
\label{eq:lkdisp}
\end{equation}
where $\Delta \mu = \hbar \omega_c + \sqrt{\pi/2} (e^2/\ell)$ is the
chemical potential discontinuity which is shifted upward by
electron-electron interactions.  The electron and hole in a magnetic field
moving parallel to each other with a velocity just sufficient so that
their oppositely directed Lorentz forces cancel their attractive Coulombic
forces. Requiring the group velocity in Eq.~(\ref{eq:lkdisp}) to give
canceling Lorentz and Coulomb forces implies that the electron and hole in
$|\Psi_{\pol{k}}\rangle$ to be separated by $ \hat z \times \pol{k}
\ell^2 $; this is consistent with the excitonic correction to the
particle-hole excitation energy in Eq.~(\ref{eq:lkdisp}). In fact this
property of $|\Psi_{\pol{k}}\rangle$ can be established explicitly. We
should expect this picture to break down when
$|k| \ell \sim 1 $ since the shortest possible localization length which
is $\sim \ell$ becomes comparable to the separation between the particle
and the hole. In fact it turns out that for small $|k|$
\begin{equation}
E(k) = \hbar \omega_c + e^2 k /2
\label{eq:skdisp}
\end{equation}
which is the long wavelength dispersion for the classical magnetoplasmon
collective mode at long wavelengths. It seems that the neutral
excitations gradually crossover from having a collective character to
having a particle-hole character as the wavevector increases. This
contrasts with the familiar case at zero magnetic field where collective
excitations and particle-hole excitations are clearly distinct, although
the fact that particle-hole excitations capture more and more oscillator
strength at larger wavevectors is still reflected in the strong field
behavior. By analogy with what happens at $\nu =1$ we expect that the
single-mode-approximation magnetoroton dispersion should be accurate for
$k \ell $ small. This has been confirmed~\cite{mred,haldanebook} by
comparisons between the single-mode-approximation theory and numerical
exact diagonalization calculations. Results of some of these calculations
are shown in Fig.~[\ref{fig:mr}].

\section{Hierarchy States}

So far we have discussed only the case of the fractional quantum Hall
effect at filling factor $\nu = 1/m$ where
$m$ is an odd integer. Experimentally the fractional quantum Hall effect
occurs at a large number of other filling factors. Up to the present
time the fractional quantum Hall effect has been observed, for fully
spin polarized electron systems at
$\nu = \nu_n \equiv n/(2n+1)$ and for $\nu = 1 - \nu_n$ for
$n=1,2,3,4,5,6,7$ as well as for $\nu = 1/5,4/5,2/7,5/7,2/9$. The largest
charge gaps occur for $\nu = 1/3$ and $\nu = 2/3$. We have argued that
these experimental results tell us that the interacting two-dimensional
electron system has chemical potential discontinuities at all these
filling factors. Presumably there are other chemical potential
discontinuities which are still masked by disorder even in the highest
quality samples available today and the ones we see have the largest
charge gaps. We would like to be able to estimate the charge gaps and to
understand why it is at these particular filling factors that the
fractional quantum Hall effect occurs; of particular interest is the
prominence of the series of filling factors $\nu_n$ and$1- \nu_n$. This
series of fractions shows up prominently in numerical exact
diagonalization calculations~\cite{gros} for the hard-core model as shown
in Fig.~[\ref{fig:hcnum}].
For reasons which will become clear it is common to refer to the
incompressible ground states at filling factors $\nu \ne 1/m$ as hierarchy
states. The reader is warned that we are now approaching a border on the
map of knowledge in the fractional Hall regime. {\em This way lie
dragons.} We start this section with some exact results and, with that
equipment in hand, venture into more rugged territory.

\subsection{Bosonization}
We now discuss a method of bosonization of the many-fermion system in a
strong magnetic field which is different from, but not unrelated to, the
statistical transmutation transformation discussed earlier. It is useful
to start with an observation concerning the many-fermion wavefunction for
a full Landau level. For finite $N$, a full Landau level state is formed
by occupying all single-particle states from
$m = 0$ to $m = N-1$. The many-fermion wavefunction is a single Slater
determinant of the form
$\Psi [z] = P[z] \prod_k \exp{(-\vert z_k\vert^2/4)}$ where
\begin{equation}
 P[z] = \left\vert\begin{array}{ccc}
z_1^0 & \ldots & z_N^0\\
\vdots & & \vdots\\
z_1^{N-1} & & z_N^{N-1}\end{array}\right\vert = \prod_{i<j} (z_i - z_j) \equiv
P_V[z].
\label{eq:vdm}
\end{equation}
The equality in Eq.~(\ref{eq:vdm}) can be understood by noting that
$P[z]$ must change sign when any pair of particles are interchanged and
therefore must vanish when $z_i = z_j$ for any $i \ne j$. It follows
that $z_i - z_j$ is a factor of $P[z]$ for any $i \ne j$.
$P_V[z]$ is the lowest degree polynomial with this property and one can
easily check that it has the same degree ($M= N (N-1)/2$) as the Slater
determinant. This proves the equality up to a constant; we leave the
proof that the constant is equal to one as an exercise. More generally any
wavefunction in the lowest Landau level must be of the same form and must
be antisymmetric. It follows that we can always write
\begin{equation}
P[z] = Q[z] P_V[z]
\end{equation}
where $Q[z]$ is a homogeneous {\em symmetric} polynomial and therefore is
the polynomial part of a {\em many-boson} wavefunction. Multiplying a
many-boson wavefunctions by $P_V[z]$ increases its homogeneous degree by
$N (N-1)/2$. It follows from the fact that multiplying polynomials is
equivalent to adding areas that if $Q[z]$ represents a boson fluid with
filling factor $\nu_B$ then $P[z]$ represents a fermion fluid with filling
factor satisfying
$\nu_F^{-1} = \nu_B^{-1} + 1$.

This mapping between a many-fermion system in the fractional Hall regime
at filling factor $\nu$ and a many-boson system at filling factor $\nu_B
= \nu / (1 - \nu) $ can be made more precise. We consider a many-fermion
system with $N$ electrons which we allow to occupy orbitals with
$m = 0, \cdots, N_{\phi}-1$ so that the filling factor is $\nu = N/
N_{\phi}$. The set of
\begin{equation}
g_{N,N_{\phi}} = { N_{\phi}! \over N! (N_{\phi} - N)! }
\end{equation}
many-fermion states corresponds to the set of independent antisymmetric
polynomials in $N$ coordinates for which the maximum power of any
coordinate is $N_{\phi}-1$. Each of these antisymmetric polynomials is
divisible by $P_V[z]$ with a quotient which is a symmetric polynomial.
The maximum degree to which any coordinate can appear in the symmetric
polynomial is $N_{\phi}- 1- N \equiv N_{\phi}^B -1$, {\em i.e.}, all
quotients belong to the Hilbert space of $N$ bosons in the fractional
Hall regime which are in a Landau level containing $N_{\phi}^B$ orbitals.
(The number of independent symmetric polynomials in $N$ coordinates with
maximum power $k$ is equal to the number of independent antisymmetric
polynomials with maximum power $k+N$. It is easy to perform the division
by $P_V[z]$ explicitly~\cite{macd85} for small numbers of electrons and
small maximum powers.) Matrix elements of any operator expressible in
terms of fermion coordinates, and in particular of the Hamiltonian, will
be preserved if they are transformed according to
\begin{equation}
\hat{O}_B[z] = \bar{P}_V[z] \hat{O}_F P_V[z].
\end{equation}
(Note that the many-boson inner product is defined by
$\hat {O}_F = 1$.) We will make arguments below based on the assumption
that these changes do not change the physics in any essential way.

It is useful togive some simple examples of boson wavefunctions and
their fermion counterparts. The lowest order symmetric polynomial has
degree $M^B =1$;
\begin{equation}
Q[z] = \sum_i z_i
\end{equation}
$Q[z] P_V[z]$ is an antisymmetric polynomial of degree $N(N-1)/2 +1$.
There is only one such polynomial;
\begin{equation}
P[z] = \left\vert\begin{array}{ccc}
z_1^0 & \ldots & z_N^0\\
\vdots & &\\
z_n^{N-2} & \ldots & z_N^{N-2}\\
z_1^N & & z_N^N\end{array}\right\vert \\.
\end{equation}
We see that the boson state in which $N-1$ bosons are in the state with
angular momentum $0$ and one boson is in the state with angular momentum
$1$ is equivalent to the fermion state in which a particle-hole excitation
has been made at the edge of the full Landau level state by promoting a
single electron from angular momentum $N-1$ to angular momentum $N$.
Another example was already introduced in connection with Laughlin's
quasihole state, $Q[z] = \prod_i z_i$. This polynomial has $M_B = N$ and
increases the power of every coordinate by one in every term of $P_V[z]$
so that
\begin{equation}
P[z] = Q[z] P_V[z] = \left\vert\begin{array}{ccc}
z_1^1 & \ldots & z_N^1\\
\vdots & &\vdots\\
z_1^N & & z_N^N\end{array}\right\vert.
\end{equation}
The boson state where all $N$ bosons have angular momentum $1$
corresponds to a fermion state in which the $m=0$ state is empty. This
is a state with a single integer charge hole in a full Landau level. As
we've discussed previously when this $Q[z]$ multiplies the $\nu =1/m$
Laughlin state it creates a state with a fractionally charged quasihole.

\subsection{Particle-hole symmetry}

The two-dimensional electron system in the fractional Hall regime has an
exact particle-hole symmetry whose importance in constructing a theory of
hierarchy states was first emphasized by Girvin~\cite{girvhole}. The
existence of this symmetry is most easily established by working in an
occupation number representation where the Hamiltonian is
\begin{equation}
H = \frac{1}{2} \sum_{m_1,m'_1,m_2,m'_2} e^{\dagger}_{m'_1}
e^{\dagger}_{m'_2} e_{m_2} e_{m_1} \langle m'_1, m'_2 | V_{ee} |
m_1, m_2 \rangle .
\label{eq:honr}
\end{equation}
Here $e^{\dagger}_m$ and $e_m$ are the creation and annihilation operators
for the state with angular momentum $m$ and we have used a standard
notation for two-body matrix elements of the electron electron
interaction. We make the particle-hole transformation by defining
creation and annihilation operators for holes using,
\begin{eqnarray}
h_m^{\dagger} & \equiv & e_m \\
h_m & \equiv & e_m^{\dagger}.
\label{eq:pht}
\end{eqnarray}
The holes operators obey the same anticommutation operators as electron
operators. The vacuum state for electrons is related to the vacuum state
for holes by
\begin{equation}
| 0 \rangle_e = \prod_m h_m^{\dagger} | 0 \rangle_h.
\end{equation}
We wish to show that the Hamiltonian for holes is identical to that for
electrons up to a constant which is proportional to the difference
between the number of electrons and the number of holes. To see this we
normal order the two body piece in the Hamiltonian expressed in terms of
hole creation and annihilation operators. This procedure leaves us with
some one body terms which we need to evaluate;
\begin{eqnarray}
e^{\dagger}_{m'_1} e^{\dagger}_{m'_2} e_{m_2} e_{m_1}
&=& h_{m_1}^{\dagger} h_{m_2}^{\dagger} h_{m'_2} h_{m'_1}\nonumber\\
&&- \delta_{m_1,m'_2}h_{m'_1} h^{\dagger}_{m_2}
 + \delta_{m_1,m'_1} h_{m'_2} h^{\dagger}_{m_2} \nonumber\\
&&- \delta_{m'_2,m_2} h^{\dagger}_{m_1} h_{m'_1}
 + \delta_{m'_1,m_2} h^{\dagger}_{m_1} h_{m'_2}.
\label{eq:hno}
\end{eqnarray}
The one-body terms turn out to be constants. To see this it is useful to
write the two-body matrix elements in terms of the Fourier representation
of the electron-electron interaction:
\begin{eqnarray}
\langle m'_1, m'_2 | V_{ee} | m_1, m_2 \rangle &=&
\int \frac{d^2 \pol{k}}{(2 \pi)^2 } V_{ee}(\pol{k}) e^{- |k|^2
}\nonumber\\
&&\times G_{m'_1,m_1}( k) G_{m'_2,m_2}(-k).
\label{eq:tbme}
\end{eqnarray}
Eq.~(\ref{eq:tbme}) follows from the expression for plane-wave matrix
elements in the lowest Landau level, Eq.~(\ref{eq:pwme}). We can now use
other identities we derived from the algebra of our ladder operator
solution of free particle problem. {}From Eq.~(\ref{eq:fll}) it follows
that
\begin{eqnarray}
\sum_{m_1} \langle m_1 m'_2 | V_{ee} | m_1, m_2 \rangle &=&
\delta_{m'_2,m_2} (2 \pi \ell^2 )^{-1} V_{ee}(\pol{k}=0)\nonumber\\
&\equiv& \delta_{m'_2,m_2} 2 \epsilon_H
\label{eq:hartree}
\end{eqnarray}
where $\epsilon_H$ is the Hartree energy per electron of the ground state
when the Landau level is full. Similarly from
Eq.~(\ref{eq:matrixproduct}) it follows that
\begin{eqnarray}
\sum_{m_1} \langle m'_1 m_{1} | V_{ee} | m_1, m_2 \rangle &=&
\delta_{m'_1,m_2} \int \frac{d^2 \pol{k}}{(2 \pi)^2}
V_{ee}(\pol{k}) e^{- |k|^2/2}\nonumber\\
&\equiv& \delta_{m'_1,m_2} 2 \epsilon_X
\label{eq:exchange}
\end{eqnarray}
where $-\epsilon_X$ is the exchange energy per electron of the ground
state when the Landau level is full. For the two models featured in these
notes $\epsilon_H$ can be calculated analytically. For the Coulomb model
$\epsilon_H$ is zero because of the omnipresent neutralizing background
required to get a finite zero of energy. For the hard-core model it
follows from Eq.~(\ref{eq:vqvm}) that $\epsilon_H = V_1$. Similarly,
exchange energies of the full Landau level are
$ \sqrt{\pi/8} (e^2/ \ell)$ and $ - V_1$ for the Coulomb and hard-core
models respectively. Combining these results the Hamiltonian can be
rewritten in terms of normal ordered hole creation and annihilation
operators;
\begin{eqnarray}
H &=& \frac{1}{2} \sum_{{m_1 , m_2}\atop {m'_1, m'_2}} h^{\dagger}_{m'_1}
h^{\dagger}_{m'_2} h_{m_2} h_{m_1} \langle m_{1}, m_{2} | V_{ee} |
m'_1, m'_2 \rangle\nonumber \\
&&+ (\epsilon_H - \epsilon_X) (N - N_h)
\label{eq:honrholes}
\end{eqnarray}
where $N_h \equiv N_{\phi} - N$ is the number of holes in the Landau
level.

Since the matrix elements of the electron-electron interaction terms are
real it follows that the spectrum of $H$ for a given number of electrons
in a Landau level is identical to that for a given number of holes in a
Landau level apart from a known constant which can be interpreted as the
interaction of the holes with its vacuum (which is a full Landau
level). In particular, the excitation spectra are identical at filling
factors $\nu$ and $1-\nu$. The energy per electron, $\epsilon$, in the
ground state satisfies,
\begin{equation}
\nu \epsilon ( \nu ) = (1 - \nu ) \epsilon ( 1 - \nu) + (2 \nu -1 ) (
\epsilon_H - \epsilon_X)
\label{eq:phse}
\end{equation}
and the chemical potential $\mu ( \nu)$ satisfies
\begin{equation}
\mu( \nu ) + \mu ( 1 - \nu) = 2 (\epsilon_H - \epsilon_X).
\label{eq:phsmu}
\end{equation}
The chemical potential discontinuities at filling factors
$\nu$ and $1-\nu$ are identical. (I leave it as an exercise for the
reader to determine how the correlation functions at filling factors $\nu$
and $1 - \nu$ are related.)

Bosonization and particle-hole symmetry allow a given situation to be
described in languages which appear to be somewhat different. Consider,
for example, the case where the Landau level filling factor is near
one. It may be more economical, for example, to use particle-hole
symmetry to describe the system as consisting of $N_h = N_{\phi} - N$
Fermi particles rather than $N$ Fermi particles. In both descriptions the
number of single-particle states available to theFermi particles is
$N_{\phi} = N + N_h$. As shown above we can do this without approximation.
Now we can apply the bosonization transformation. The number of states
available to the Bose particles is then reduced by the number of
particles {\em i.e.}, from $N + N_h$ to $N$. The bosonization
transformation is not as clean since it requires a change in the
Hamiltonian and the inner product which we don't know how to treat
exactly. We assume on reasonable physical grounds that this change is
unimportant; there is fairly strong evidence~\cite{fbtrans} to support
this assumption from numerical exact diagonalization transformations.
Note that the number of states available to these Bose particles is $N$;
it is as if the Bose particles lie in a Landau level created by a magnetic
field of one flux quantum for each electron. We'll see this result again
soon, coming from a different but related direction.

\subsection{Correlation factors}

There is another non-unitary transformation of the many-electron
wavefunctions which is physically important:
\begin{equation}
\Psi'[z] = \prod_{i<j} (z_i - z_j)^2 \Psi [z].
\label{eq:cfac}
\end{equation}
According to our angular-momentum-counting arguments this transformation
changes the filling factor according to
\begin{equation}
\nu'^{-1} = 2 + \nu^{-1}.
\label{eq:ffcfac}
\end{equation}
I will refer to this as the `correlation-factor' transformation. For $\nu
> 1/3$ this transformation maps the Hilbert space at filling factor $\nu$
to the part of the Hilbert space at filling factor $\nu^{'}$ which is
projected out by letting $V_1 \to \infty$. For sufficiently
short-range interactions this transformation should therefore map the
many-particle Hilbert space at filling factor $\nu$ to the low-energy
portion of the many-particle Hilbert space at filling factor $\nu'$. It
is common~\cite{macd85} to argue on physical grounds that the changes in
the Hamiltonian and the inner product which result from the
transformation are sufficiently innocuous that the transformation maps
eigenstates to eigenstates, although the numerical evidence is less
convincing in this case~\cite{morfcf}. However, the weaker conclusion that
for sufficiently short-range interactions an incompressibility at filling
factor $\nu$, implies an incompressibility at filling factor $\nu^{'}$ is
well supported both by numerical calculations and by the ultimate jury,
experiment.

It is possible to `derive' the Laughlin states using the
correlation-factor transformation. Starting from the full Landau level
state at $\nu =1$ the correlation factor transformation generates a
sequence of non-degenerate grounds states at $\nu = 1/m$ for odd $m$
which are precisely the Laughlin states. The transformation also maps
states with $\nu < 1$ to states with $\nu < 1/3$. The reader can verify
that the state with a single hole in a full Landau level, discussed
above, is mapped precisely to Laughlin's approximate wavefunction for a
single fractionally charged quasihole in the incompressible $\nu = 1/3$
state. Similarly states with many holes are mapped to states with many
fractionally charged quasiholes. Since the number of particles and the
dimension of the Hilbert space are preserved by the correlation-function
transformation it follows that if the quasiholes are regarded as fermions
they have Landau levels with degeneracy $N+N_h$, while if they are
regarded as bosons they have Landau levels with degeneracy $N$. Systems
with many holes can form incompressible states. We know, for example,
from the particle-hole transformation that when $N_{h} = N_{\phi}/3$ the
holes form a $\nu = 1/3$ incompressible state. The correlation-factor
transformation will generate from this state an incompressible ground
state at $\nu =2/7$. Thus we can explain many of the incompressibilities
which are observed experimentally. However, no combination of the above
transformations will generate an incompressibility in the filling factor
range
$ 1/3 < \nu < 2/3$. It is precisely in this range of filling factors that
the most robust fractional quantum Hall effects occur. To account for
these incompressibilities we have a choice of stepping onto one of a
number of thinner theoretical limbs.

\subsection{Classical Hierarchies}

The original hierarchy schemes were motivated by the indication from
early experiments that the fractional quantum Hall effect for
spin-polarized electrons could occur at any rational value of $\nu = q/p$
with $p$ odd. Related pictures which could account for an
incompressibility at all such filling factors were advanced initially by
Laughlin~\cite{lauh}, Haldane~\cite{pseudopots}, and
Halperin~\cite{halpqpstat}. Attempts to obtain more quantitative
estimates of hierarchy state properties were also
made~\cite{macd85,macdplasma} using similarly motivated approximations
These three pictures all start from the observation that the Laughlin
incompressible states have fractionally charged quasiparticles and attempt
to describe the system at nearby filling factors in terms of quasiparticle
degrees of freedom. The three pictures differ in the statistics assumed
for the quasiparticles. As recognized by Halperin~\cite{halpqpstat}, the
three pictures are distinguished only if predictions depend on the
statistics representation in which long-range gauge forces, representing
fluctuations about an average magnetic field, are absent. It turns out
that this distinction does not lead to any difference in the filling
factors at which incompressible states are expected.

We describe this classical hierarchy pictures by following Haldane and
assuming that the quasiparticles are bosons. We start by considering
filling factors $\nu < 1/m$ so that the system contains a dilute gas of
charge $1/m$ quasiholes. As we've discussed above, the low-energy part of
the Hilbert space is that of $N_h$ bose particles in a Landau level with
degeneracy $N$. The low-energy part of the spectrum will be broadened by
interactions between the quasiholes and we expect that the ground state
will be incompressible when the quasiholes can form a boson Laughlin state
{\em i.e.}, when $N_h/N = 1/2n$. Since for this case $N_{\phi}=m N +
N_{h}$ this happens when the electron filling factor is $(m+1/2n)^{-1}$.
We get precisely the same result when we describe the quasiholes as
fermions since the particles are then in a Landau level with degeneracy
$N+N_h$. So far everything is based on the relatively solid results of
the previous subsections. The hierarchy picture is based on the
expectation that the same ideas should apply to fractionally charged
quasielectron excitations which are created when $N_{\phi}$ is decreased
at fixed $N$. In the quasielectron case Bose Laughlin states for the
quasielectrons occur when $N_{\phi} = m N - N/2n$ or at filling factor $(m
- 1/2n)^{-1}$. Although we cannot offer compelling analytical arguments
in favor of treating the quasielectrons as a Bose gas in a Landau level
with degeneracy $N$, we do have a substantial~\cite{hiernum} body of
detailed numerical studies which support this notion. The line of argument
can be continued by assuming that for filling factors near
$(m \pm 1/2n)^{-1}$ we can apply the same argument to the quasiparticles
of the Bose Laughlin states formed by the quasiparticles at the first
level of the hierarchy. It is possible to demonstrate that all filling
factors with an odd denominator can be generated in this way.

\subsection{Neoclassical Hierarchies}

A picture of the hierarchy states can be generated by generalizing the
Chern-Simons-Landau-Ginzburg picture of the fractional quantum Hall
effect~\cite{cslghier}. In this theory the quasiparticles are associated
with vortices of the Bose superfluid. The hierarchy is constructed by
taking advantage of the approximate particle-vortex duality of
superfluids. The gas of quasiparticles, which was mapped to a gas of
vortices in a superfluid by the Chern-Simons-Landau-Ginzburg theory is
mapped back to a gas of Bose particles using the particle-vortex duality
of superfluids.  These Bose particles can then have a fractional quantum
Hall effect along the lines of the original hierarchy.

A separate picture of the hierarchy which is associated with a fermion to
fermion statistical transmutation arose from several related lines of
investigation. A system of electrons at filling factor $p / (2np \pm 1)$
has $ 2 n \pm 1/p $ flux quanta for each electron. We can apply a
statistical transmutation in which $2n$ flux-quanta, directed in
opposition to the flux from the physical magnetic field,  are attached
to each electron. When this flux is treated in a mean-field
approximation the result is a system with $p$ electrons for each quantum
of net magnetic flux {\em i.e.,} the mean field system has $p$ filled
Landau levels. If the mean-field approximation is justified, the
fractional quantum Hall effect could then be explained as an integer
quantum Hall effect of composite fermions~\cite{jaincf} consisting of
electrons and an even number of attached flux quanta. It can be argued
that the neglect of fluctuations is justified by the gap between Landau
levels associated which produces the integer quantum Hall effect. This is
not sufficient, however, since fluctuations evidently alter the physics
completely when there are no interactions between the electrons and the
fractional quantum Hall effect does not occur. The simplest application
of this approach is to the Laughlin filling factors which are generated
from this approach with $p=1$ where it is the fermion representation
equivalent of the Chern-Simons-Landau-Ginzburg theory. Lopez and
Fredkin~\cite{lopezfradkin} have shown that the Laughlin wavefunctions can
be derived in this approach when the fluctuations are treated in a random
phase approximation. At this level accounting for the fluctuating
magnetic fields associated with $2m$ flux quanta on the particles give
rise to the correlation factors, $\prod (z_i -z_j)^{2m}$ whose
significance was explained from a microscopic point of view above.

For $p > 1$ this transformation maps the system at the mean field level to
a system with more than one filled Landau level. One appealing aspect of
this approach is that the sequence of fractions generated with the
smallest values of $m$ are precisely the set of fractions for which the
most robust fractional quantum Hall effects occur experimentally. At
least at a superficial level this provides a rationale for this
experimental fact. A difficulty with this approach becomes more apparent
when we consider $p > 1$. The mean-field state has components in higher
Landau levels which are unphysical in the strong magnetic field limit.
As we have mentioned earlier any approach based on statistical
transmutation requires that we enlarge our usual Hilbert space to include
the high energy states with electrons in higher Landau levels. In the
composite fermion approach the gap at the mean-field level is proportional
to $\hbar \omega_c$, a quantity which has absolutely no meaning in the
fractional Hall regime. If it were possible to treat the fluctuations
accurately this scale would be replaced by the interaction scale $e^2 /
\ell$, but to date it has only been possible to due this
phenomenologically. Despite this weakness, explicitly trail wavefunctions
proposed by Jain~\cite{jaincf}, in which correlation factors are attached
to the mean-field states and a projection onto the lowest Landau level is
performed, have proved to be remarkably accurate for system containing a
small number of electrons. For example Jain has proposed the following
wavefunction,
\begin{equation}
\Psi^{J}_{2/5}[z] = \overline{\prod_{i<j} (z_i - z_j)^2 \Psi_{\nu=2}[z]}
\end{equation}
where the overbar implies projection onto the lowest Landau level, as a
trial wavefunction for the incompressible ground state at $\nu = 2/5$.

\subsection{Overview on Hierarchy States}

We have seen that there are several approaches for understanding
hierarchy states which appear on the surface to be quite different. All
are phenomenological in that they are unable to predict quantitative
values for measurable quantities, such as the charge gaps at particular
filling factors. They are therefore distinguishable on the basis of
qualitative predictions and here, as shown by Read~\cite{readhp}, there
are no differences. Every approach makes identical predictions for the
excitation structure at every hierarchy state filling factor. All
approaches can be generalized so that they can generate fractional quantum
Hall effects at all filling factors with odd denominators.  There
are, however, some mild differences in the weak statements that can be made
about
the values of $\nu$ at which the fractional quantum Hall effect occurs.
In one common interpretation, the classical hierarchies predict that if,
in a given sample, the fractional quantum Hall effect does not occur at a
particular filling factor, then it cannot occur at any filling factor
which evolves from that filling factor at a later generation in the
hierarchy. I am not aware of violations of this prediction in experiment.
On the other hand the composite fermion approach has the advantage that
the filling factors in the `main sequence' ($\nu = \nu_n$ or $\nu_n = 1 =
\nu_n$) appear on a more equal footing.

The composite fermion approach also makes interesting predictions for the
properties of the system at filling factors for which the ground state is
not incompressible and the fractional quantum Hall effect does not
occur. For $\nu = 1/2n $, there exist fermion to fermion statistical
transmutation transformations in which the mean-field state is a fermion
gas in no magnetic field. In this case, the many electron system should
be a Fermi liquid if the fluctuations in the statistical magnetic field
are sufficiently innocuous. Halperin, Read, and Lee have recently
completed a thorough theoretical study of this scenario~\cite{hlr} which
analyses the role of electron-electron interactions in suppressing
fluctuations at low order in perturbation theory. On the experimental
side several recent studies~\cite{halfexp} have uncovered very suggestive
evidence of Fermi-liquid-like behavior for $\nu$ near $1/2$. Experiment
provides even clearer evidence of an important length scale in the system
which appears to diverge like the cyclotron orbit radius associated with
the mean magnetic field seen by composite fermions as $\nu$ approaches
$1/2$.

\section{What's not here}

These notes provide an introduction to physics in the quantum Hall
regime. This continues to be a very active area of physics both
theoretically and experimentally and there are many fascinating topics
which we have not been able to even touch upon. I conclude by mentioning
a few of these.

In the integer quantum Hall regime theory~\cite{critsigxx,chalker}
predicts that at the energy where the localization length diverges
$\sigma_{xx} = \sigma_{xy} = 0.5 e^2 / h$. So far it is not very clear
that these expectations are consistent with experiment. One possible
explanation is that even weak interactions change the value of
$\sigma_{xx}$ at the critical energy. The whole area of the interplay
between disorder and interactions has not been thoroughly explored. It
appears~\cite{yang} that interactions create a dip in the tunneling
density-of-states even in the integer quantum Hall regime. Weak
interactions influence physical properties in the integer quantum Hall
regime in a way which interpolates, as the localization length changes,
between being similar to interaction
effects in insulators and being similar to interaction effects in metals.

Gapless edge excitations are an
essential companion of the quantum Hall effect as we have emphasized. The
structure of the edge excitation spectrum in the fractional case can be
quite intricate~\cite{macdedge,wenedge} in reflection of the complicated
nature of the hierarchical ground states. The edge system comprises a
chiral realization of a one-dimensional electron gas which is
predicted~\cite{wenedge,fk} to have the power law behavior of
low-temperature physical properties associated with the breakdown of
Fermi liquid theory in interacting one-dimensional electron systems. The
unique aspect of fractional Hall systems as one dimensional electron
systems is that the exponents may be related exactly to the dimensionless
quantized Hall conductance rather than being non-universal quantities
dependent on high-energy physics. Experimental study of these effects
has recently been initiated.\cite{webb}

Finally I would like to mention recent work on
double-layer systems~\cite{doublelayer} which can exhibit a number of
highly unusual properties associated with unusual broken symmetry in which
phase coherence occurs between electrons in different layers which are
isolated apart from having inter-layer Coulomb interactions.
Experimental study of these phenomena has also been initiated
recently and some interesting phenomena have been observed.\cite{dlexpt}

\acknowledgments

The scope and depth of these notes has been limited both by the scope and
depth of my understanding of the field and by the estimate of your
patience made by the editors of this volume in limiting the length of this
contribution.  A number of interesting topics have not been touched upon
at all.  For example I have discussed neither the crossover between
integer and fractional quantum Hall effects~\cite{kivelson} as a function
of disorder strength nor the crystallization transition which occurs at
small filling factors. Nevertheless I hope that you have found these pages
helpful. I share the credit for anything admirable with those who have
shared their insights on the quantum Hall effect with me, especially Luis
Brey, Marcus Buttiker, Ren\'{e} C\^{o}t\'{e}, Jim Eisenstein, Matthew
Fisher, Herb Fertig, Duncan Haldane, Michael Johnson, Catherine Kallin,
Klaus von Klitzing, Bob Laughlin, Rudolf Morf, Phil Platzman, Mark
Rasolt, Ed Rezayi, Mansour Shayegan, Shivaji Sondhi, Pavel Streda, David
Thouless, Eric Yang, Daijiro Yoshioka,  Xiao-Gen Wen, Shou-Cheng Zhang and
the person in the adjacent office, Steve Girvin.  I would like to thank
Erik S{\o}rensen for a careful reading of a draft of these notes.  I
apologize for obscurities and deficiencies in scholarship and encourage
you to tactfully bring them to my attention.  Support from the US National
Science Foundation under grant DMR-9416906 is gratefully acknowledged.

\begin{figure}
\caption{Band edge profiles for a typical two-dimensional electron gas
system at a $GaAs$ $Al_xGa_{1-x}As$ interface. The electrons are bound
to the interface because of charge transfer of electrons from ionized
donors in the larger bandgap semiconductor. I will limit my attention
here to the case where the two-dimensional electron gas occurs in the
conduction band, as illustrated. I take the direction perpendicular to
the interface to be the $\hat z$ direction. Its single-particle
Schr\"{o}dinger equation has bound states at the interface whose
energetic separation may be considered as a large energy. This leads to
two-dimensional `subbands'; we will assume that all electrons are confined
to the lowest energy `subband'. Electrons move freely parallel to the
interface but see a random potential which has contributions from remote
ionized donors, as well as from ionized acceptors in the $GaAs$ and from
imperfections in the interface between the two-semiconductors. The
effective mass for the two-dimensional electron system in the lowest
subband depends weakly on the details of a particular system but is
typically quite close to the bulk $GaAs$ conduction band effective mass,
$m^* \approx 0.067 m_e$.}
\label{fig:2deg}
\end{figure}

\begin{figure}
\caption{Hall bar geometry. In this typical six probe measurement a
current, $I$, flows from source to drain. The dissipative resistance,
$R$, is the ratio of the voltage drop along the direction of current flow
($V_A-V_B$ or $V_D-V_C$) to $I$. The Hall resistance, $R_H$, is the
ratio of the voltage drop across the sample ($ V_A - V_D$ or $V_B-V_C$ )
to $I$.}
\label{fig:hallbar}
\end{figure}

\begin{figure}
\caption{Dissipative and Hall resistivity data for a typical
two-dimensional electron gas system in the integer quantum Hall regime.
These measurements were made by H.P. Wei and D.C. Tsui.}
\label{fig:intqhe}
\end{figure}

\begin{figure}
\caption{Dissipative and Hall resistivity data for a typical
two-dimensional electron gas system in the fractional quantum Hall
regime. These measurements were made by H.P. Wei and D.C. Tsui.}
\label{fig:fracqhe}
\end{figure}

\begin{figure}
\caption{Complex number description of a classical cyclotron
orbit.}
\label{fig:cyorb}
\end{figure}

\begin{figure}
\caption{A large but finite two-dimensional electron gas. In panel (a)
the chemical potential lies in a gap and the only low-energy excitations
are localized at the edge of the system. In panel (b) the chemical
potential lies in a mobility gap so that there are low-energy excitations
in the bulk but they are localized away from the edge. In panel (c) a net
current is carried from source to drain by having local equilibria at
different chemical potentials on upper and lower edges.}
\label{fig:incomp}
\end{figure}

\begin{figure}
\caption{The rate at which the edge current changes with chemical
potential in a gap (a) and in a mobility gap (b). This illustration is for
the case where the disorder potential varies in one direction only. The
properties at the edge do not change as the scale of the disorder is
increased so that states localized in the bulk of the system occur at the
chemical potential.}
\label{fig:mobility}
\end{figure}

\begin{figure}
\caption{Illustration of the Laughlin's gedanken experiment leading to
fractionally charged quasiparticle states when an incompressibility occurs
at fractional Landau level filling factors}
\label{fig:fraccharge}
\end{figure}

\begin{figure}
\caption{Schematic illustration of `flux-tube' attachment.}
\label{fig:fluxattach}
\end{figure}

\begin{figure}
\caption{Magnetoroton excitation energies for $\nu =1/3$, and $\nu = 1/5$
calculated by Girvin, MacDonald, and Platzman.}
\label{fig:mr}
\end{figure}

\begin{figure}
\caption{Finite size estimates of the dependence of the chemical
potential on filling factor for the hard-core model.
$N_L$ is the degeneracy of the lowest Landau level on the sphere and the
chemical potential is estimated from
$\nu \approx \mu_n \equiv E_0(N)-E_0(N-1)$.}
\label{fig:hcnum}
\end{figure}


\begin{references}
\bibitem{disorderpot} Some quantitative issues will depend on the precise
nature of the disorder potential.
\bibitem{velocity} Corresponding definitions apply for the velocity ladder
operators which change the Landau level indices.
\bibitem{caveatflux} A word of caution is in order. The effect of a
magnetic flux on the system appears through the kinetic energy term which
becomes irrelevant when we project onto a particular Landau level. In
order to consider the effect of such a flux we need to relax the
projection onto a single Landau level, perform the operation, and if
appropriate reapply the projection.
\bibitem{caveatpa} The zero of the potential is arbitrary and linear terms
could also in principle be present without contributing to the
Laplacian. The fact that linear terms are absent means that the force due
to the background potential vanishes at the origin from which we can
deduce that the background charge which produces the potential is
symmetric around the origin. We can imagine the background charge as a
disk of uniform plasma charge density whose radius is larger than the
maximum radius at which electrons can be found with significant
probability. The electron system will then be centered on the origin.
\end{references}
\end{document}